\documentclass[11pt,a4paper]{article}
\usepackage{jheppub}

\usepackage{graphicx}
\usepackage{dcolumn}
\usepackage{bm}

\usepackage{booktabs}
\usepackage{amsmath,amssymb,amsbsy,amstext, amsthm, amsfonts}
\usepackage{graphicx}
\usepackage{color}
\usepackage{slashed}
\usepackage[dvipsnames]{xcolor}
\usepackage{tikz}
\usepackage{dsfont}
\usepackage{verbatim}
\usepackage[utf8]{inputenc} 
\usepackage{ragged2e}

\newcommand{\half}{\frac{1}{2}}

\newcommand{\gen}{{N_{g}}}
\renewcommand{\t}[1]{\tilde{#1}}
\newcommand{\bb}[1]{\mathbb{#1}}

\newcommand{\mpl}{M_\text{pl}}
\def\Vhrulefill{\leavevmode\leaders\hrule height 0.7ex depth \dimexpr0.4pt-0.7ex\hfill\kern0pt}

\newcommand\extrafootertext[1]{%
    \bgroup
    \renewcommand\thefootnote{\fnsymbol{footnote}}%
    \renewcommand\thempfootnote{\fnsymbol{mpfootnote}}%
    \footnotetext[0]{#1}%
    \egroup
}

\def\@fnsymbol#1{\ensuremath{\ifcase#1\or $\Re$\or $\Im$\or  \else\@ctrerr\fi}}
\renewcommand{\thefootnote}{\fnsymbol{footnote}}

\begin{document}

\title{\vspace{-0.5cm}A Cosmological Lithium Solution from Discrete Gauged Baryon Minus Lepton Number}

\author{Seth Koren}
\affiliation{Enrico Fermi Institute, University of Chicago,
Chicago, IL 60637, U.S.A. \\
Oehme Postdoctoral Fellow, sethk@uchicago.edu, they/him
}

\abstract{
The cosmological lithium problem---that theory predicts a primordial abundance far higher than the observed value---has resisted decades of attempts by cosmologists, nuclear physicists, and astronomers alike to root out systematics. 
We reconsider this problem in the setting of the Standard Model extended by gauged baryon minus lepton number, which we spontaneously break by a scalar with charge $2 N_c N_g$. 
Cosmic strings from this breaking can support interactions converting three protons into three positrons, and we argue that an `electric'-`magnetic' interplay can give this process an amplified, strong-scale cross-section in an analogue of the Callan-Rubakov effect.
We suggest such cosmic strings have disintegrated $\mathcal{O}(1)$ of the primordial lithium nuclei, and lay out what is necessary for this scheme to succeed. 
To our knowledge this is the first new physics mechanism with microphysical justification for the abundance of lithium uniquely to be modified after Big Bang Nucleosynthesis. \\


{\centering \vspace{-15pt}
\includegraphics[height=8cm,width=\textwidth,angle=180]{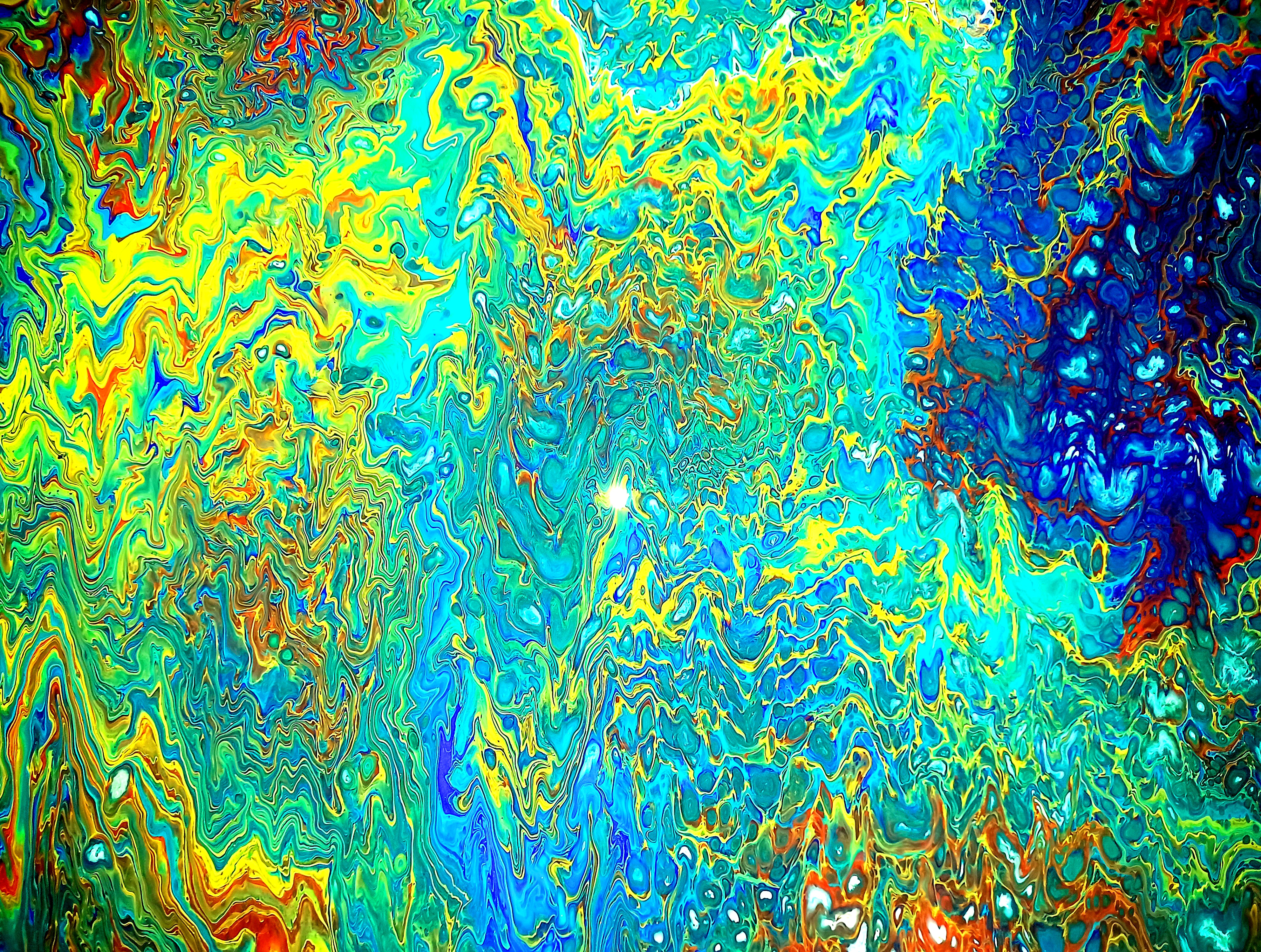}\\ \vspace{+10pt}
\footnotesize Figure 0: Photo of a commissioned abstract artist's impression in acrylic on canvas by Bailey Lingen, a Chicago-based artist and trans rights activist who produces music under the name LiTHiUM THiEF.
}
}

\maketitle

\tableofcontents 
\vspace{-1cm}\noindent\textcolor{white}{\rule{2cm}{1cm}}

\section{The Lithium Problem}

\begin{table}[t]
    \centering
    \begin{tabular}{|c|c|} \hline
    & $\left({}^7\text{Li}/\text{H}\right)$ \\ \hline
    Observation & $\left( 1.6 \pm 0.3 \right) \times 10^{-10}$   \\ \hline
    Theory & $\left(4.7 \pm 0.7\right)\times 10^{-10}$   \\ \hline
    \end{tabular}
    \caption{For reference, the status of the number density of lithium-7 relative to that of hydrogen, taken from the recent review and status update \cite{Fields:2019pfx}. \vspace{-0.5cm}}
    \label{tab:lithium}
\end{table}

Big Bang Nucleosynthesis (BBN) is the theory of the formation of light nuclei a few minutes after the hot big bang, and is the earliest epoch of the universe for which we have direct evidence. In the framework of the Standard Models (SM) of particle physics and cosmology, BBN is essentially a parameter-free cross-check on our understanding of the universe. 
And a powerful one at that: BBN is sensitive to all four known forces, as it pits the strong force's ability to bind nucleons against the gravitational expansion of the universe, the weak decays of the neutron, and the electromagnetic repulsion of protons. 
The status is given by the comparison of theoretical predictions and empirical observations of just three quantities \cite{Fields:2019pfx,Iocco:2009a,Coc:2017pxv, Mathews:2017xht}: The primordial abundances of deuterium with one proton, of helium-4 with two protons, and of lithium-7 with three protons. It's these three values that characterize our only semi-direct understanding of the universe far before recombination. 

While theory and observation agree well for hydrogen and helium, there has been a consistent mystery regarding the abundance of lithium \cite{Fields:2011,Iocco:2012a} beginning with observations in 1982 by Spite and Spite \cite{Spite:1982a,Spite:1982b}, and the current best values are shown in Table \ref{tab:lithium}.
The concern, of course, is what sort of systematic effects might be responsible.  
Early on there was a possibility of unseen effects in the nuclear data \cite{Cyburt:2009cf}, but the relevant transition rates are thought to have now all been measured, with by 2015 a status update stating ``a `nuclear option' to the lithium problem is essentially excluded'' \cite{Cyburt:2008kw,Cyburt:2015mya,Iliadis:2020a}. Continued measurement and modelling has agreed \cite{Pitrou:2018cgg,Rijal:2019a,nTOF:2018vnr,Bystritsky:2017a} and ever-more-inventive channels are being studied by nuclear physicists and discarded as inconsequential \cite{Voronchev:2009zz,Kirsebom:2011vp,Li:2018a,Hartos:2018tpy,Damone:2018a}. 

The remaining worry arises since the empirical data on `primordial' lithium comes primarily from measurements of its abundance 
in the atmospheres of metal-poor, population II stars in the Milky Way. While we must necessarily use nearby sources due to the scarcity of lithium, 
the argument that we can access the primordial abundance follows from elementary stellar dynamics considerations and was given by Spite \& Spite \cite{Spite:1982a,Spite:1982b}.  
In relatively cool stars, the full volume of the star undergoes convection, such that lithium from the atmosphere may be convected to the core and destroyed. However, hot-enough stars develop an inner radiative zone which is convectively stable, such that convection is confined to an outer layer and the lithium abundance in the atmosphere does not get reprocessed. 
Then if one measured the lithium abundance in atmospheres of stars whose lithium is primordial, as a function of temperature---with all of the proper astronomical care taken to look at the right sorts of stars---for cooler stars there should be less and less of it, but on the other side at some temperature the plot should level off to a constant value. Indeed, this is the famous `Spite plateau' which was observed first by Spite and Spite in 1982 \cite{Spite:1982a,Spite:1982b}, suggesting that this plateau abundance indeed grants us an observational window to the primordial abundance.
The next few decades of observational progress (e.g. \cite{Rebolo:1988a, Balachandran:1990a,Ryan:1996a,Ryan:1998a,Ryan:1999a,Boesgaard:2005a,Charbonnel:2005a,Bonifacio:2006au,Asplund:2006a,Aoki:2009a,Hosford:2009a,Sbordone:2010a,Schaeuble:2012a}) in measuring stellar lithium are reviewed in \cite{Spite:2012}, overall showing further support for a Spite plateau discrepant with the BBN predictions. 
This has now included observations of the Spite plateau in environments as diverse as stellar streams \cite{Aguado:2021a}, multiple globular clusters \cite{Bonifacio:2002a,Mucciarelli:2011a,Schiappacasse-Ulloa:2022}, the Sculptor dwarf spheroidal galaxy \cite{Hill:2019a}, the Gaia-Enceladus galaxy \cite{Molaro:2020a,Simpson:2021a}, and Messier 54 \cite{Mucciarelli:2014a}. A recent study of galactic chemical evolution models for these data concluded that the observed abundances were well-understood only if the primordial lithium abundance was set to the Spite plateau value \cite{Matteucci:2021}.

In recent years some ever-larger datasets have showed slightly higher variability \cite{Bonifacio:2006au,Bonifacio:2007a,Gonzalez:2009a,Monaco:2012a,Crandall:2014cca,Melendez:2010a,Howk:2012a} in the pleateau star abundances and even some anomalously low abundances in extremely metal-poor stars \cite{Sbordone:2010a,Bonifacio:2015top}, which has raised doubts in some quarters (see e.g. \cite{Deal:2021a}), but the implications of these new data are unclear \cite{Aguado:2019bac,Matteucci:2021} and 
the situation is still under active investigation by many scientists in a wide variety of fields across physics and astronomy. 
On the theoretical side, there is no well-accepted new stellar dynamics mechanism that could be responsible, though there have been proposals \cite{Richard:2005a,Piau:2006a,Korn:2006a,Prantzos:2012a,Ouyed:2013a,Marrder:2020a,Tognelli:2020a,Grisoni:2019a}. On the observational side, astronomers continue to further probe the Spite plateau in a variety of innovative directions \cite{Prodanovic:2004a,Ruchti:2011a,Caffau:2011a,Mucciarelli:2012,Hansen:2014a,Spite:2015,Fu:2018a,Bensby:2018a,Lambert:2004a,Pinto:2021a}.\footnote{It is also important to note when reading earlier literature that claims about positive detections of lithium-6 have now been called into question \cite{Fields:2022mpw}.}

Of course, any time there is an astronomical or cosmological anomaly our Bayesian priors should be peaked toward learning something new about stars. But 
after so long with the prospects for an astrophysical resolution still unclear, some 
broader exploration is surely well worth doing. And indeed, there have been
some new physics mechanisms explored---mainly fundamental constants varying to modify early-universe nuclear rates \cite{Dent:2007a,Dmitriev:2004a,Flambaum:2008a,Berengut:2010a,Cheoun:2011yn,Landau:2006a,Coc:2007a,Mori:2019a,Franchino-Vinas:2021nsf,Clara:2020efx,Fung:2021wbz} or decaying dark matter disrupting reactions \cite{Pospelov:2006sc,Jedamzik:2007cp,Ellis:2005ii,Olive:2012hig,Coc:2014gia,Yamazaki:2017uvc,Cyburt:2006a,Kaplinghat:2006a,Bird:2008a,Kohri:2007a,Kusakabe:2010a,Kusakabe:2011a,Anchordoqui:2020djl,Alcaniz:2019kah,Sato:2016len,Cyburt:2013fda,Goudelis:2015wpa,Pospelov:2010cw,Bailly:2008yy,Cumberbatch:2007me,Jedamzik:2005dh,Cyburt:2012kp,Jittoh:2011ni,Kawasaki:2010yh,Kusakabe:2017brd,Kusakabe:2014ola}. See also \cite{Mathews:2019hbi,Luo:2019a,Makki:2019a,Malaney:1993a,Mathews:2005a,Hou:2017a,Nakamura:2017a,Lu:2022a,Flambaum:2018ohm,Pospelov:2012pa,Bailly:2010hh,Cyburt:2010vz,Olive:2012xf,Coc:2013a,Pospelov:2010hj} for further ideas. In our diagnosis, these approaches have been stymied by not uniquely picking out lithium to be affected, and they generally encounter severe difficulties with putting other observables into tension. 

We are motivated by the surprisingly long lifetime of the proton and its exact stability in the SM to take advantage of the exact SM selection rule \begin{align*}\hspace{1cm} \Delta(\# \text{ of baryons}) &= \Delta(\# \text{ of leptons}) \\ &= 0 \ (\text{mod number of generations}).\end{align*} Since the SM has $N_g = 3$ generations of particles and the BBN predictions first fail for lithium with three protons, this symmetry could explain why elements with one or two protons are unaffected.
This allows us to construct---to our knowledge---the first new physics mechanism for the lithium abundance to be discrepant which does not require tuning to avoid modifying the abundances of other elements. 


We must go only slightly beyond the SM to find our effect of interest.
In brief, we will break gauged baryon minus lepton number to a discrete subgroup which will produce a large abundance of cosmic string loops. These topological defects amplify the $B+ N_c L$ violation, and we introduce a model in which leptoquarks communicate this violation in a way which picks out solely processes in which three protons are turned into three positrons. We propose that such cosmic strings have disintegrated lithium nuclei in the early universe, reducing the primordial abundance to the level inferred from astronomical observations. In this \textit{Letter} we will put together the various pieces of field theory and cosmology behind this mechanism, and estimate the rate to demonstrate its plausibility. Precise predictions will require much further work.

\section{Schema}\label{sec:body}

\begin{table}[t]\centering
\normalsize
\begin{tabular}{|c|c|c|c|c|c|c|}  \hline
 & $\quad Q \quad$ & $\quad\bar u\quad$ & $\quad\bar d\quad$ & $\quad L \quad$ & $\quad\bar e\quad$ & $\quad \bar \nu \quad$ \\ \hline
$SU(3)_C$ & $3$ & $\bar 3$ & $\bar 3$ & -- & -- & --\\ \hline
$SU(2)_L$ & $2$ & -- & -- & $2$ & -- & --\\ \hline
$U(1)_{Y}$ & $+1$ & $-4$ & $+2$ & $-3$ & $+6$ & --\\ \hline
$U(1)_{B}$ & $+1$ & $-1$ & $-1$ & -- & -- & --\\ \hline
$U(1)_{L}$ & -- & -- & -- & $+1$ & $-1$ & $-1$\\ \hline 
\end{tabular}\caption{Representations of the left-handed SM Weyl fermions (and the sterile neutrino) under the classical symmetries of the SM. }\label{tab:charges}
\end{table}

\paragraph{New physics setting}

We extend the Standard Model by gauging $U(1)_{B- N_c L}$, which is the continuous, generation-independent, Adler-Bell-Jackiw (ABJ) \cite{Adler:1969gk,Bell:1969ts} anomaly-free  `accidental' global symmetry of the SM.
\footnote{We normalize the charges such that the least charged particles in the unbroken phase of the SM (the quarks) have unit charge, which alleviates confusion when properly counting integer invariants. After QCD confines, it is $(B-N_c L)/N_c$ that counts baryons and leptons.} The fermion charges are given in Table \ref{tab:charges}. 
We address the 't Hooft anomaly \cite{tHooft:1979rat} and neutrino oscillations by adding right-handed neutrinos, but this choice does not impact our mechanism. 
Though note that $U(1)_{B-N_c L}$ does forbid Majorana masses for neutrinos. 
To Higgs this gauge symmetry we introduce a scalar $\Phi$ with 
\begin{equation}
    \left[\Phi\right]_{B-N_c L} = 2 N_c \gen, \qquad \left[\Phi\right]_{B+N_c L} = 0, \qquad \langle \Phi \rangle = v,
\end{equation}
which condenses in the early universe at a temperature $T_c \sim v$. This breaks the baryon minus lepton number symmetry without breaking the additional discrete anomaly-free global $\bb{Z}^L_{\gen}$ subgroup of lepton number enjoyed by the SM \cite{Koren:2022bam}. 
The global symmetry-breaking pattern providing selection rules is
\begin{equation}
    U(1)_{B-N_c L} \times \bb{Z}^L_{\gen} \rightarrow \bb{Z}_{2 N_c \gen}^{B- N_cL} \times \bb{Z}^L_{\gen} \supset \bb{Z}_{2 N_c \gen}^{B+N_c L},
\end{equation}
where the anomaly-free subgroup of baryon plus lepton number is generated by $(1,1) \in \bb{Z}_{2 N_c \gen}^{B-N_c L} \times \bb{Z}^L_{\gen}$. We emphasize that only the former factor is gauged in our model, though it would be interesting to consider a model in which the discrete lepton number is gauged as well. 
While this Higgsing is `non-minimal' in the sense that $\left[\Phi\right]>1$, this choice of $\left[\Phi\right]$ preserves the SM selection rule which stabilizes the proton \cite{Tong:2017oea,Koren:2022bam,Anber:2021upc,Davighi:2022qgb,Wang:2022eag}. That is, the SM constraint of $\Delta B/N_c = \Delta L = N_g$, already familiar from electroweak sphaleron processes, will remain exact in our theory.  

\paragraph{Discrete gauge theory}

The nontrivial Higgs charge $\left[\Phi\right]_{B-N_c L}$ means its condensation preserves an unbroken discrete $\bb{Z}_{2 N_c \gen}^{B-N_c L}$ gauge symmetry \cite{Krauss:1988zc,Alford:1989ch,Preskill:1990bm,Coleman:1991ku,Banks:1989ag} and stable cosmic string solutions exist with tension $\mu \sim v^2$. The transverse slices of a static, $z$-independent string are Abrikosov-Nielsen-Olesen vortices \cite{Abrikosov:1956sx,Nielsen:1973cs} and have asymptotics
\begin{equation} \label{eqn:anovortex}
    \Phi(\rho,\theta) \rightarrow v e^{ik\theta}, \qquad  \vec{A}(\rho,\theta) \rightarrow \frac{k}{g} \frac{1}{2 \gen} \frac{\hat{\theta}}{\rho},
\end{equation}
with $\vec{A}$ the $B-N_c L$ gauge boson, $g$ the $U(1)_{B-N_c L}$ `electric' charge, and $k\in \bb{Z}$ the winding number of $\Phi$. With nonzero winding, continuity in the interior demands a zero in the Higgs field,\footnote{By applying Brouwer's fixed point theorem to $\Phi(x)$ on each transverse disk.} 
and indeed the $B-N_c L$ symmetry is unbroken in a string `core' of radius $R \sim v^{-1}$. These solutions have discrete $U(1)_{B-N_c L}$ `magnetic' fluxes 
\begin{equation}
    \Phi_B = \frac{2\pi}{g} \frac{k}{2  N_c \gen}
\end{equation} 
confined to their cores. This magnetic flux is in `fractional' units of $1/(2  N_c \gen)$ that of the Dirac magnetic flux quantum $\Phi_0 = 2\pi/g$ \cite{Dirac:1931kp}, so the string is stable and interacts via the Aharonov-Bohm (AB) effect \cite{Aharonov:1959fk,Aharonov:1984zza} for the unbroken discrete symmetry. The cosmic strings undergo elastic AB scattering with all the SM fermions of differential cross-section per unit length (for a plane wave of $B- N_c L$ charge $q$ and momentum $p$ incident on a string) \cite{Alford:1988sj}
\begin{equation} \label{eqn:discreteab}
    \frac{1}{\ell} \frac{d\sigma}{d\theta} = \frac{\sin^2(\pi q \Phi_B/\Phi_0)}{2\pi p \sin^2(\theta/2)},
\end{equation}
which, importantly for the phenomenology, 
is independent of any high scales or small couplings. \footnote{This is nicely understood by dualizing the infrared physics to the topological BF theory \cite{Horowitz:1989ng,Horowitz:1989km}, which is a generalized, covariant theory of this discrete AB effect. In this dual description, these are cosmic BF strings which are charged under an emergent two-form $\bb{Z}_{2N_c\gen}$ gauge symmetry \cite{Teitelboim:1985ya,Banks:2010zn}. The interaction between point charges and strings then must depend solely on the product of electric and magnetic charges of the $\bb{Z}_{2N_c\gen}$ BF theory, $k q \text{ (mod } 2 N_c \gen)$ \cite{Teitelboim:1985yc}.}

\paragraph{Strings in cosmology}

During an early universe phase transition, topological defects are produced proportional to the correlation length $\xi$ of the order parameter  fluctuations.
Causality ensures this is no larger than the Hubble distance $\xi \lesssim d_h$, so that at a critical temperature $T_c\sim v$, $\mathcal{O}(1)$ horizon-crossing strings of length $d_h = 2 H^{-1}(T_c) \sim \mpl/v^2 = (G \mu)^{-1} \ell_{pl}$ per Hubble volume are formed \cite{Kibble:1976sj,Kibble:1980mv}, $d_h$ being the Hubble size, $H$ the Hubble parameter, and $G\mu \sim v^2/\mpl^2$ a dimensionless measure of the gravitational effects of the strings. At formation, these long strings contain energy density 
\begin{equation}
    \rho_{string} \sim \frac{\mu d_h}{d_h^3} \sim (G\mu)^{+1} \rho_{tot},
\end{equation} a small fraction of the total. 
The elastic AB scattering with the SM plasma provides a total transverse cross-section per unit length of $\sigma_\perp/\ell \sim 1/T$, 
which leads to a long epoch of friction-dominated evolution until the temperature drops below $T_f \lesssim v^2/\mpl$ \cite{Vilenkin:1991zk}.

When cosmic strings encounter each other, they `intercommute' with probability approaching unity \cite{Manton:1981mp,Shellard:1987bv,Matzner:1988qqj,Ruback:1988ba,Shellard:1988ki,Atiyah:1985dv,Gibbons:1986df,Rosenzweig:1990ea,Samols:1991ne,Abdelwahid:1993fg,MacKenzie:1995np,Gauntlett:2000ks,Hanany:2005bc,Hashimoto:2005hi}. This results in the production of small-scale structure on the strings \cite{Blanco-Pillado:2015ana,Kibble:1990ym,Copeland:2009dk,Kibble:1990ym,Siemens:2001dx,Siemens:2002dj} and many closed loops from strings crossing themselves. 
String loops oscillate and shrink via gravitational radiation \cite{Vilenkin:1981kz,Vilenkin:1981bx,Vilenkin:2000jqa}, which dissipates energy and prevents the strings from dominating the density budget \cite{Turok:1984db}. 

After decades, we now understand the behavior of the cosmic string network at late times and on large scales: There is an attractor `scaling' solution where the length spectrum of loops depends only on the ratio with the horizon distance $\ell/d_h$ \cite{Albrecht:1984xv,Bennett:1987vf,Shellard:1987bv,Shellard:1988zx,Matzner:1988qqj,Moriarty:1988em,Myers:1991yh}.
The small scale behavior of this `scaling' population is assumed to continue down to the `gravitational cutoff' scale $\ell \sim (G \mu) d_h$, at which length a loop evaporates within a Hubble time, which leads to a large number of `small' such loops, $n \sim (G \mu)^{-0.7}$ \cite{Blanco-Pillado:2013qja}.

However, the regime relevant for their effects on lithium requires understanding cosmic strings far before the infrared fixed point is reached.
In addition to being very difficult to study, this is sensitive to details such as the strength of the phase transition, since the correlation length of Higgs fluctuations may in fact be quite small compared to the horizon length $\xi \ll d_h$, resulting in a far greater initial number density of defects \cite{Guth:1982pn,Zurek:1985qw,Murayama:2009nj}. 
The small-scale behavior is also sensitive to the details of intercommutation dynamics, and furthermore the presence of currents on the strings may stabilize them against decay and result in a large portion of energy density in cosmic string `vortons' \cite{Davis:1988ij,Brandenberger:1996zp,Davis:1996zg,Carter:1999an,Peter:2013jj,Auclair:2020wse,Battye:2021kbd}. 
For these reasons, rather than attempting to model the evolution of the phase space of cosmic strings, below we will simply parametrize their effects on lithium disintegration in terms of the fraction of the energy density which is in cosmic strings.

\begin{table}[t]\centering
\normalsize
\begin{tabular}{|c|c|c|c|c|c|}  \hline
 & $\quad \Phi \quad$ & $\quad \chi \quad$ & $\quad \omega_\ell \quad$ & $\quad \omega_q \quad$ & $\quad \omega_s \quad$ \\ \hline
$SU(3)_C$ & -- & -- & $3$ & $3$ & -- \\ \hline
$SU(2)_L$ & -- & -- & -- & -- & -- \\ \hline
$U(1)_{Y}$ & -- & -- & $-8$ & $-8$ & -- \\ \hline
$U(1)_{B}$ & $+9$ & $+9$ & $+1$ & $-2$ & $+3$ \\ \hline
$U(1)_{L}$ & $-3$ & $+3$ & $+1$ & $0$ & $+1$ \\ \hline 
\end{tabular}\caption{Representations of the scalars added in our benchmark model. 
Cosmic strings appear from the winding of $\Phi$ in the vacuum, the condensation of $\chi$ on the string allows it to freely exchange $B+ N_cL$, and the scalars $\omega$ communicate $B+N_cL$-breaking by $\chi$ to the right-handed SM fields.
}\label{tab:newcharges}
\end{table}

\paragraph{Topological defect catalysis}

We wish to consider interactions which destroy lithium nuclei by turning three protons into three positrons, thereby violating $B+N_c L$ by $2N_c \gen$ units. We utilize a stringy analogue of the Callan-Rubakov effect where grand unified monopoles `catalyze' proton decay at strong scale rates $\sigma(p^+ + \text{ monopole} \rightarrow e^+ + \text{ monopole}) \sim \Lambda_\text{QCD}^{-2}$. For monopoles, the inevitability of such inelastic interactions can be seen purely from the infrared \cite{Dirac:1931kp,Dirac:1948um,Goldhaber:1965cxe,Weinberg:1965rz,Zwanziger:1968ams,Peres:1968umt,Lipkin:1969ck,Wu:1976qk,Schwinger:1966nj,Zwanziger:1968rs,Zwanziger:1970hk,Zwanziger:1972sx,Kazama:1976fm,Callan:1983ed,Csaki:2020inw,Csaki:2020yei}. An ultraviolet description for the monopole \cite{tHooft:1974kcl,Polyakov:1974ek,Arafune:1974uy,Prasad:1975kr,Julia:1975ff,Hasenfratz:1976vb,Tomboulis:1975qt,Callan:1975yy,Boulware:1976tv} allows one to resolve the microphysical interactions and see $p^+ \rightarrow e^+$ arising from interactions with leptoquarks in the core and the anomalous violation of $B+ N_cL$ \cite{Jackiw:1975fn,Hasenfratz:1976gr,Goldhaber:1976dp,Goldhaber:1977xw,Callan:1982ac,Rubakov:1981rg,Rubakov:1982fp,Callan:1982ac,Callan:1982au,Callan:1982ah,Callan:1983tm,Dawson:1983cm,Sen:1984qe,Polchinski:1984uw,Isler:1987xn,Rubakov:1988aq,Affleck:1993np,Brennan:2021ewu,Brennan:2021ucy,Csaki:2021ozp}.

The cosmic string case has seen far less study, and no infrared argument for this catalysis effect is yet known. To provide an explicit benchmark---and to give a microphysical model targeting this mechanism toward the cosmological lithium problem---we introduce ultraviolet physics which lets us map our case on to a model studied by Alford, March-Russell, \& Wilczek in 1989 \cite{Alford:1989ie} in which a leptoquark condenses inside cosmic strings of some broken gauged $\widetilde{U(1)} \rightarrow \bb{Z}_N$. In that model the leptoquark allows $p^+ \rightarrow e^+$ violation of $B+ N_cL$, but here we respect the exact SM symmetry $\bb{Z}_\gen^L$, so rather than the condensation of a leptoquark 
we instead introduce a gauge-singlet scalar $\chi$ with
\begin{equation}
    \left[\chi\right]_{B-N_cL} = 0, \qquad \left[\chi\right]_{B+N_c L} = 2  N_c \gen , \qquad \langle \chi \rangle = 0,
\end{equation} but which condenses on the string core so long as
\begin{equation}
    \left. \partial_\chi^2 V(\Phi, \chi)\right|_{\Phi=0} < 0 \rightarrow \langle \chi \rangle_{\Phi=0} = v_\chi.
\end{equation} 
This breaks solely the global approximate $U(1)_{B+N_c L}$ on the string core down to the subgroup preserved by the SM, and we take the symmetry-breaking scales equal for simplicity, $\langle \Phi \rangle = v$ around the vacuum and $\langle \chi \rangle_{\Phi = 0} = v$ in the string. This condensate realizes a global baryon plus lepton number analogue of Witten's bosonic superconductivity \cite{Witten:1984eb}, which allows the string core to exchange $2 N_c \gen$ units of $B+N_cL$ charge with the outside world \cite{Witten:1984eb}. 

\begin{figure}[h]
    \centering
        {\includegraphics[width=0.40\textwidth]{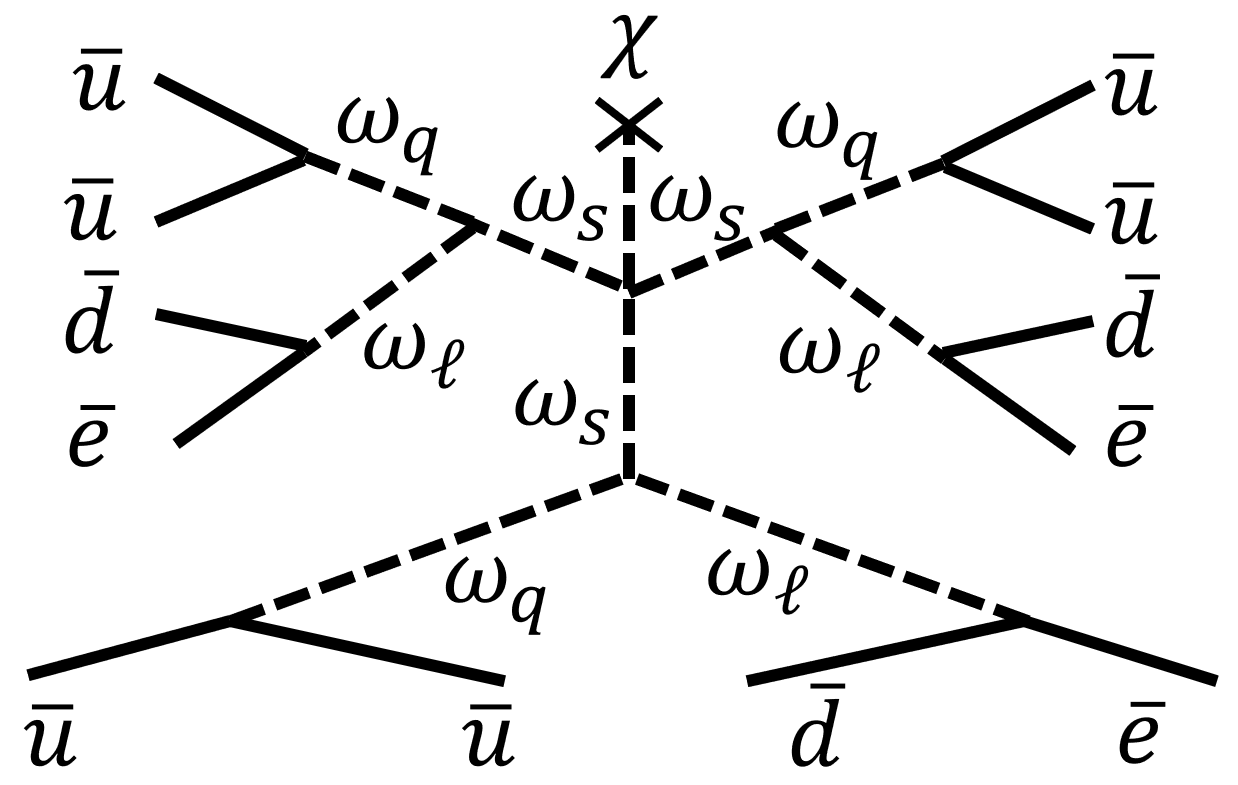}}
    \caption{A Feynman diagram portraying our benchmark for how the $B+N_c L$-breaking effects of the $\chi$ vev get communicated to the SM fermions at low energies.}
    \label{fig:feynDiag}
\end{figure}

The infrared symmetries forbid interactions of $\chi$ with fewer than twelve SM fermions, but we can explicitly generate one such interaction by introducing the scalar leptoquark $\omega_\ell$ with hypercharge $8$ and its analogue diquark $\omega_q$, as well as the gauge singlet scalar $\omega_s$ with $B=N_c L=3$. This choice of scalars (see Table \ref{tab:newcharges}) allows the nontrivial interactions 
\begin{equation}
    \mathcal{L} \supset \lambda_\ell \omega_\ell \bar d \bar e + \lambda_q \omega_q^\dagger \bar u \bar u + \kappa \omega_\ell^\dagger \omega_q \omega_s + \lambda_\chi \omega_s^3 \chi^\dagger. 
\end{equation}
Around the vacuum the Feynman diagram portrayed in Figure \ref{fig:feynDiag} generates the operator
\begin{equation} \label{eqn:feynOp}
    \mathcal{L} \sim \frac{1}{v^{15}}\mathcal{\chi}(\bar u \bar u \bar d \bar e)^3,
\end{equation} and $B+N_c L$-violation on the string core is communicated only to electrons and protons in our model. This appears naturally due to our choice of leptoquark representations, but it would be interesting to consider relaxing this and examining the possibility of interactions with helium as well. 
We further take the mediators $\omega$ to be light in the string core, so this suppression by $v$ disappears when the SM fermions are interacting with the $\chi$ vev in the core, \footnote{Microphysically, this requires quartic interactions with $\chi$, $V \propto |\chi|^2 |\omega_i|^2$ to be much smaller than those with $\Phi$, $V \propto |\Phi|^2 |\omega_i|^2$. Our intention here is just to evince some model in which the effect of interest occurs, but one interesting model-building direction is to sharpen this microphysical model---perhaps with unbroken supersymmetry in the string core which can leave these scalars light.} 
and we've set all couplings to unity and scales equal for simplicity.

In the model of Alford, March-Russell, \& Wilczek \cite{Alford:1989ie}, the Yukawa interaction allows an incoming quark to inelastically scatter into a lepton by exciting the leptoquark condensate, and the $\bb{Z}_N$ discrete gauge field background enhances the cross-section. When the discrete gauge charge of the incoming quark is maximal, the cross-section is unsuppressed by ultraviolet scales, as in the Callan-Rubakov effect. In our case, we instead have  
an interaction of many SM partons with the string condensate. 

However, in the cosmological case of interest, a ${}^7\text{Li}$ nucleus delivers a state of three protons $\psi_{3p} \sim (\bar u \bar u \bar d)^3$ incident upon the string. To map our model to \cite{Alford:1989ie}, we rewrite Equation \ref{eqn:feynOp} as a Yukawa interaction between $\psi_{3p}$, the nine quarks in three protons, and $\psi_{3e}$, the three positrons. Adapting their results while ignoring many subtleties about this matching, we infer that an incoming plane wave of $B-N_c L$-charge $q$ and momentum $p$ undergoes enhanced inelastic scattering with a long cosmic string as 
\begin{equation}
    \frac{\sigma}{\ell} \sim \frac{1}{p}\sin^2\left(\pi \alpha\right) \left(\frac{p}{v}\right)^{4\left|\alpha - \half\right|}, \qquad \alpha \equiv \frac{kq}{2 N_c \gen},
\end{equation}
where we are generalizing Eqns 18-21 of \cite{Alford:1989ie}. The inelastic scattering is unsuppressed by ultraviolet scales for an incoming state with $B-N_c L$ charge $q= N_c \gen$ in the case of a single-winding string. We note also that topological defect interactions can violate crossing symmetry, as may be understood simply for monopoles \cite{Csaki:2020inw}, so for example crossing one proton to an outgoing antiproton results in a drastically suppressed cross-section because the incoming state no longer has maximal $\mathbb{Z}_{2N_g N_c}$ charge nor half-integer spin, which is also crucial for catalysis \cite{Alford:1989ie}. 
This suppression is useful for ensuring consistency with observations of primordial helium, and it would be interesting to understand precisely the subleading rate at which helium might be affected.

\paragraph{Parameter space}

In the early universe after BBN, just such a $q =   N_c \gen$ state of three protons may be delivered to the string within the nuclear potential well of ${}^7\text{Li}$.\footnote{In fact BBN initially produces ${}^7\text{Be}$ which then captures an electron to produce ${}^7\text{Li}$  \cite{Khatri:2010ed}. While in much of the parameter space it is ${}^7\text{Be}$ which is being destroyed, in any case these are the only elements with more than 2 protons, so our selection rule still serves to distinguish them.} We assume string loops have a cross-section per unit length
\begin{equation} \label{eqn:xsecBenchmark}
    \frac{\sigma}{\ell}\left( 3 p^+ + \text{ string} \rightarrow 3 e^+ + \text{ string} \right) \sim \Lambda_\text{QCD}^{-1},
\end{equation}
but note again that there are necessarily large theoretical uncertainties here.

\begin{figure}[t]
    \centering
        {\includegraphics[clip, trim=0.0cm 0.0cm 0.0cm 0.0cm, width=0.5\textwidth]{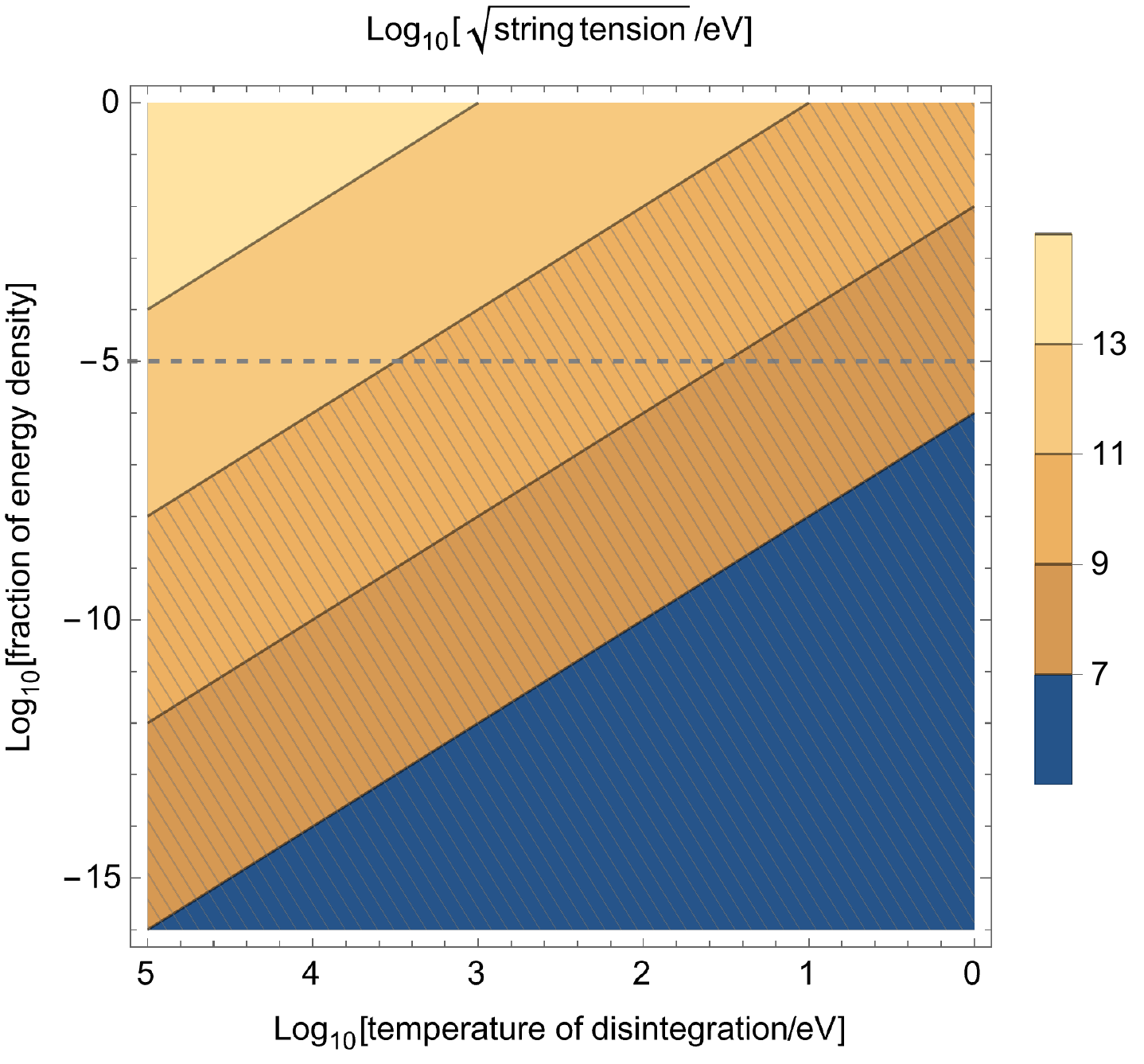}}
    \caption{A contour plot of the string tension scale required for $\mathcal{O}(1)$ lithium to be disintegrated as a function of the temperature and the fraction of energy density in cosmic strings from Equation \ref{eqn:fracVeVTemp}. 
    The hatched region is constrained in simple models of gauged $B-N_c L$ with a massive $Z'$, taking $\sqrt{\mu} \sim v \sim m_{Z'}/g > 100 \text{ GeV}$ as a benchmark \cite{Heeck:2014zfa}.
    The dashed line is an approximate upper bound on the energy density in strings near the Hubble length from CMB and gravitational wave constraints \cite{Sanidas:2012ee,Rybak:2021scp}.
    }
    \label{fig:fracVevTemp}
\end{figure}

However a first question is just whether such interactions can plausibly destroy an order-one fraction of lithium nuclei after BBN. Due to the complications discussed above, we will not attempt to model the evolution of the phase space of string loops, but instead content ourselves to parametrize their effects as a function of the fraction of energy density in cosmic strings. We simply ask that the rate for a lithium nucleus to encounter a cosmic string be greater than Hubble,
\begin{equation}
    \Gamma \simeq \int d\ell \frac{dn}{d\ell} \sigma \simeq \Lambda_\text{QCD}^{-1} \int d\ell \frac{dn}{d\ell} \ell =  \frac{\rho_{\text{string}}}{\Lambda_\text{QCD} \mu}\gtrsim H,
\end{equation}
where we have begun with a general differential number density of cosmic strings $\frac{dn}{d\ell}$, and assumed the interaction is characterized as $\sigma/\ell = \Lambda_\text{QCD}^{-1}$ which makes the rate proportional to the energy density. Then introducing $f \equiv \rho_{\rm string}/\rho_{\rm tot}$ we require 
\begin{equation}\label{eqn:fracVeVTemp}
    f \gtrsim \frac{\Lambda_{\rm QCD}}{M_{\rm pl}} \frac{\mu}{T_\star^2},
\end{equation}
where $\Lambda_{\rm QCD}/M_{\rm pl} \sim 10^{-20}$, and $T_\star$ is the temperature at which disintegration occurs. We note that this is a greater fraction than that eventually expected in long strings $f \sim G \mu$, underscoring the necessity of understanding the nonequilibrium behavior of strings after the phase transition. Along similar lines, we emphasize also that the precise requirement to destroy the correct fraction of lithium will depend on the duration over which these interactions are efficient. We plot this relation in Figure \ref{fig:fracVevTemp}, where we assume that $\mu \gtrsim T_{\rm BBN}^2$ such that the $B-N_c L$ phase transition takes place before BBN, and $T_{\rm CMB} \lesssim T_\star \lesssim T_{\rm BBN}$ to avoid severe constraints on exotic energy injection which are sensitive even down to the level of $\mathcal{O}(10^{-10})$  at the time of recombination \cite{Poulin:2016nat,Acharya:2019uba,Dienes:2018yoq,Slatyer:2016qyl,Acharya:2019owx}. If the $B- N_c L$-breaking scale is low then one requires relatively small gauge coupling \cite{Heeck:2014zfa,Ilten:2018crw,Bauer:2018onh}, but the `electric'-`magnetic' interplay of the Callan-Rubakov effect means that this does not suppress the cross-section, as above. Nevertheless if the scale is so low that the string width is larger than nuclear scales, the benchmark cross-section, which does not account for the string's internal structure, seems less applicable. 
In Figure \ref{fig:fracVevTemp} we also show a benchmark constraint on gauged $B- N_c L$ with massive $Z'$ from \cite{Heeck:2014zfa}, assuming the tension goes as $\sqrt{\mu} \sim m_Z'/g$, though we note that this constraint does assume a particular right-handed neutrino sector and it is possible these constraints are modified either with non-standard neutrino interactions (see e.g. \cite{Miranda:2015dra}) or exotic neutrino sectors (see e.g. \cite{Wang:2020mra,Wang:2020xyo,Wang:2020gqr}).

\paragraph{Complementary signatures}

Cosmic strings may generally be searched for through their gravitational effects either in affecting the CMB anisotropies or in contributing to the stochastic gravitational wave background. However, the details of the spectra and the ensuing constraints depend strongly on the size distribution of strings. To guide the eye, we offer in Figure \ref{fig:fracVevTemp} the rough constraint applicable to the scaling population of strings. For discussions on how the constraints depend on some characteristics of the spectrum we refer to \cite{Caldwell:1991jj,Sanidas:2012ee,Blanco-Pillado:2013qja,Blasi:2020mfx,Chang:2021afa,Blanco-Pillado:2021ygr} and \cite{Contaldi:1998mx,Lazanu:2014eya,Lazanu:2014xxa,Lizarraga:2016onn,Charnock:2016nzm,Rybak:2021scp}.

The non-trivial dynamics of cosmic strings means there may be a non-negligible number density at any time, so one may hope to constrain their existence with direct detection experiments. 
General $B-N_c L$ cosmic strings may be probed through their unitarity-limited elastic Aharonov-Bohm scattering, but this requires a detailed understanding of the kinematics of such interactions in direct detection experiments. The inelastic effect is more striking, but one must also understand the cross-section for general nuclei, such as the many oxygen in Super-Kamiokande.
Experimentally, dedicated analysis or new searches may be required for the `hard' signature of three outgoing positrons, their ensuing showers, and possible delayed decays of the leftover nucleus. 

While such effects are too rare with solely the scaling population of large cosmic strings, in scenarios with increased number density such as with vorton end-states this would be an interesting signature to understand. 
Alternatively it would be interesting to integrate over cosmic volumes and look for indirect effects of $B-N_c L$ strings in astrophysical settings. 
Regardless of their role in producing the lithium discrepancy, such strings have enormous elastic scattering cross-sections with SM matter from the AB effect. 

\section{Conclusion} 

In this work we have stirred together a m\'{e}lange of ingredients from particle theory and cosmology to write down a mechanism through which the lithium problem is an entr\'{e}e to fundamental physics beyond the SM. 
This scenario motivates various important directions of inquiry, which are interesting more generally than in their application here:
\begin{itemize}
    \item It is clearly important to better conceptually understand the cosmic string analogue of the Callan-Rubakov effect, which likely requires understanding BF theory with light fermions.
    \item Any precise rate predictions will require understanding the evolution and interactions of small cosmic string loops, which is a difficult problem.
    \item It would be useful to think about other interesting applications for unitarity-limited interactions of fermions with cosmic strings---perhaps in effecting baryogenesis or dark matter production.
    \item For any such applications it will be necessary to understand better how to translate from many-parton interactions with cosmic strings to nuclear effects.
\end{itemize}

Theoretical physicists have been captivated by the interplay of electric and magnetic effects for centuries now, from Maxwell \cite{Maxwell:1865zz} up to its grandest form in Montonen-Olive duality \cite{Montonen:1977sn,Goddard:1976qe,Osborn:1979tq} with maximal supersymmetry. The case of \textit{less} symmetry, as with our discrete $\bb{Z}_N$, has received far less attention. 
In addition to their fascinating topological character, these interactions may have interesting, heretofore-underappreciated phenomenological applications in the early universe. 

\section{Dedication and Acknowledgements}

I am grateful to many colleagues for relevant conversations, including Wen Han Chiu, Clay Cordova, Brian Fields, Christina Gao, Andrew Long, Ryan Plestid, Carlos Wagner, Liantao Wang, and Yiming Zhong, and I thank especially T. Daniel Brennan and Sungwoo Hong for discussion and many useful comments. I further thank Samuel Alipour-fard and Sungwoo Hong for helpful comments on earlier drafts of this manuscript, and an anonymous referee for pointing out an inconsistency among assumptions in a prior draft. I would like to also thank Bailey Lingen for their artistic creativity and enthusiasm.

This work was supported by an Oehme Postdoctoral Fellowship from the Enrico Fermi Institute at the University of Chicago, and I am sincerely grateful for the freedom this has granted me to think in offbeat directions. I also thank organizers at the Galileo Galilei Institute, Stony Brook University, the University of Chicago, and the Aspen Center for Physics who allowed me to present this work in progress, and the participants therein for interesting questions. 

\noindent \Vhrulefill\includegraphics[width=1em]{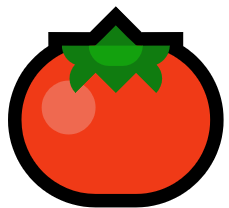}\Vhrulefill

This paper is dedicated to my mother, whose death during its writing was premature and preventable. At cause is my parents' mistaken understanding of the nature of reality, as encapsulated by the fact that they have dedicated their lives to promoting anti-vaccine misinformation. After I understood how incorrect their worldview was, in protest explicitly thereof I cut off all contact with them back in 2015.

Nonetheless, it is also the case that my mother meant well, and even my parents' considerable medical abuse was doled out with loving intentions. Whatever substantial damage this did me, it is responsible also for my need to understand how the universe works and what is possible within it. 
And \textit{if} I have here newly envisaged some interesting connection which has been lurking amongst thousands of papers since before my birth, it is surely \textit{because} of my experiences having had neuroendocrine tumors rather than in spite of that brain injury and the many related struggles.


I write in the hopes that my parents' sincere motivations,  transmuted as they have been toward my contributions to our understanding of reality, may morally counterbalance their misguided actions.


\appendix

\section{Theory Motivation}

Following the theoretical success of the SM half a century ago, brilliant and inventive physicists turned their attention and curiosity to higher energies.  They discovered remarkably interesting field theories and constructed beautiful UV pictures suggesting how the universe should work. Maybe an orbifold compactification provides super-gauge-Higgs-unification, and everything we know is brought together at high energies. Maybe supergravity mediates SUSY breaking from a dark sector. Maybe we get right-handed neutrinos to fill out our spinor $SO(10)$ representation, maybe dark matter is the lightest superpartner, maybe leptons are the fourth color of quarks. And flavor this, and CP that, and left-right so-and-so. Everything fit together as so many puzzle pieces.

So where is all of the new physics that should be there? Where are the $Z$ primes, the flavons, the leptoquarks? The KK partners, the superpartners, the mirror fermions? Where is the electron EDM, flavor-changing neutral currents, neutrinoless double beta decay, dark matter annihilation or custodial symmetry violation? And why haven't we observed proton decay? 

With each additional, amazingly precise, null finding from our experimental colleagues, it becomes more clear that something is not quite right with our grand picture. The question then becomes which cherished principles to hold fast, and which to give up. Perhaps weak scale SUSY giving $SU(5)$ gauge coupling unification is merely accidental. Maybe understanding the dimensionful parameters in the SM requires `UV/IR mixing' effects past the paradigm of local effective field theory (e.g. \cite{Koren:2020biu,Craig:2018yvw,Craig:2019fdy,Craig:2019zbn}). Perhaps tiny Dirac neutrino masses shouldn't concern us. 

Or maybe there's some hint we've overlooked, or should give greater weight. Perhaps it's of fundamental importance that we live near $\Omega_m = \Omega_{\Lambda}$, or that $\Lambda \sim m_\nu^4$ allows the existence of an $\text{AdS}_3 \times S_1$ minimum in the SM \cite{Arkani-Hamed:2007ryu}, or that supersymmetry in three dimensions does not require Bose-Fermi degeneracy \cite{Witten:1995rz}, or that we have $\gen=3$ generations of fermions. We do not yet know whence a new understanding may arise. The silver lining of this nebulous outlook is the opportunity it provides to explore novel scenarios and apply fun field theory ideas.

In this work we explore an interesting consequence of second-guessing an implicit assumption of most new physics models which we have recently emphasized in \cite{Koren:2022bam}. Most of these grand pictures explicitly break an underappreciated $\bb{Z}_{2 N_c \gen}$ subgroup of baryon plus lepton number which is part of the exact global symmetry group of the SM. In contrast, we shall make especial efforts to ensure it remains a good symmetry.

\section{Discrete Gauge Symmetries} 
\label{sec:zngen}

In this section we recall the anomaly-free, generation-independent global symmetries of the SM, before considering upgrading a subset to discrete gauge symmetries in the IR. We then discuss finding such a discrete gauge theory from the Higgsing of a continuous gauge symmetry. 
As we have emphasized in a companion note \cite{Koren:2022bam}, the SM anomaly-free global symmetries include not only the familiar continuous baryon minus lepton number $U(1)_{B-N_cL}$ but also a discrete $\bb{Z}_{2 N_c \gen}$ subgroup of baryon plus lepton number. This exact symmetry is responsible for ensuring that the proton cannot decay in the SM \cite{Byakti:2017igi,Tong:2017oea}.

If we go beyond the marginal operators composed of SM fields, we can look for higher-dimensional operators in the SM EFT which do not break the SM symmetries. Just imposing the continuous exact symmetry $U(1)_{B-N_cL}$ still allows $B+N_cL$-violating proton decay at dimension six, as commonly seen in grand unified theories. However, if one imposes \textit{all} the global symmetries of the quantum SM, the lowest-dimension operators which break $B+N_cL$ are instead
\begin{equation}\label{eqn:effop}
    \mathcal{L} \sim \frac{1}{\Lambda^{14}} (\bar u \bar u \bar d \bar e)^3
\end{equation}
at dimension eighteen, and similar operators involving neutrons and neutrinos, as well as other generations. The point being that the SM structure in itself already privileges processes in which three baryons turn into three leptons, but any such operator effect in the SM is ridiculously suppressed and certainly negligible \cite{Helset:2021plg}. We must go beyond the SM to find our effect of interest, but in fact not too far.

\subsection{Gauging SM Accidental Symmetries} \label{sec:smacc}

We now wish to consider the possibility that some of these SM global symmetries are in fact gauge symmetries at low energies. In \cite{Koren:2022bam} we have discussed that the above symmetries have no `ABJ-type' \cite{Adler:1969gk,Bell:1969ts} global-gauge-gauge anomalies, meaning that the SM gauge fields do not violate the global $U(1)_{B-N_cL}\times \bb{Z}_\gen^L$. However, to consistently gauge these we must also check for the existence of a 't Hooft anomaly \cite{tHooft:1979rat}, namely the cubic anomaly $U(1)^3_{B-N_cL}$. 

If we hope to integrate this into a theory of quantum gravity, we should consider also the mixed anomaly with gravity, $\text{grav}^2 \times U(1)_{B-N_cL}$, which just sums over all $B-N_cL$-charged degrees of freedom since gravity couples universally. In the SM both of these are nonzero
\begin{align}\label{eqn:thooftanom}
    \sum B-N_cL &= -N_g N_c, & \sum (B-N_cL)^3 &= -N_g N_c^3,
\end{align}
which is often interpreted as due to the lack of sterile neutrinos, which indeed would cancel both identically. 

So in the SM by itself, we may not gauge $U(1)_{B-N_cL}$---restricting to the SM means only the subgroup $\bb{Z}_{N_c N_g}^{B-N_cL} \times \bb{Z}^L_\gen$ may be consistently gauged. We are within our rights as effective field theorists to simply extend the SM by the discrete gauge theory of a gauge boson $A$ which satisfies $\exp(i \oint A) \in \bb{Z}_{N_c N_g}^{B-N_cL} \times \bb{Z}^L_\gen$ \cite{Tachikawa:2017gyf}. This theory SM$^+$ has the same local physics as the SM, but disallows even otherwise-conjecturally-inevitable quantum gravitational decays of the proton, since it imposes $\Delta B = N_c \Delta L = N_c N_g$ nonperturbatively. So there is some sense in which better lower bounds on the lifetime of the proton provide strictly positive Bayesian evidence toward preferring SM$^+$ over SM at low energies. But I digress.

Here we wish to begin with the full continuous $B-N_cL$ symmetry gauged in the ultraviolet, so we have two options for how to deal with the nonzero 't Hooft anomalies of Equation \ref{eqn:thooftanom}. It is possible that there are light right-handed neutrinos, meaning that our counting above is wrong as they must be included in our infrared SM. Then there is no barrier to gauging the full $U(1)$, and indeed we take this as the benchmark to keep in the back of our minds. 

Alternatively, the degrees of freedom which cancelled that anomaly in the UV theory may have gained a large mass during the $B-N_cL$ symmetry breaking $U(1) \rightarrow \bb{Z}_{2 N_c N_g}$. Once this is upgraded to gauge theory, due to the low-energy anomaly the heavy $B-N_cL$ degrees of freedom do not fully decouple. Instead, they have effects in the infrared which are universally captured by the construction of Wess \& Zumino \cite{Wess:1971yu} and Witten \cite{Witten:1983tw}. While this would generally seem to be less-UV-sensitive, it poses a challenge to standard neutrino mass mechanisms.\footnote{We note here some recent work on exotic ways of addressing this anomaly \cite{Wang:2020mra,Wang:2020xyo,Wang:2020gqr,Wang:2021ayd}.} In any case the details will not be relevant for us here, so we postpone further exploration to future model-building work. 

\subsection{$N$-Charged Abelian Higgs} \label{sec:nabelianhiggs}

We wish to study the Higgsing of $U(1)_{B-N_cL}$ down to $\bb{Z}_{2 N_c N_g}^{B-N_c L}$, and have the $\bb{Z}_\gen^L$ factor as an unbroken global symmetry.
For aesthetic reasons we discuss here a general $U(1)$ Abelian Higgs and a scalar $\Phi$ with some non-minimal charge $N>1$. The Lagrangian is
\begin{equation}
    \mathcal{L} = - \frac{1}{4} F^{\mu\nu} F_{\mu\nu} - \left|(\partial_\mu - i e N A_\mu) \Phi \right|^2 - V(\Phi),
\end{equation}
where the potential is minimized for some nontrivial vev $\langle\left|  \Phi \right|\rangle  = v$. At low energies, we expand the Higgs field around this vev in terms of a radial mode $\rho$ and a Goldstone mode $\phi$ as 
\begin{equation}
    \Phi(x) = \frac{1}{\sqrt{2}} (v + \rho(x)) e^{i \phi(x)}.
\end{equation}
In terms of these fields, a gauge transformation $A_\mu \rightarrow A_\mu + \partial_\mu \alpha(x)$ then effects a nonlinear change in the Goldstone mode $\phi \rightarrow \phi + N \alpha(x)$. 

As a result of the factor $N$, a subgroup of gauge transformations satisfying $e^{i N \alpha(x)}=1$ leaves the Higgs field unchanged. Since $\Phi$ is not charged under these symmetries, they remain unbroken when $\Phi$ condenses. While this \textit{locally} requires trivially constant gauge transformations $\alpha(x) \equiv 2 \pi k/N$ for $k=0,\dots,N-1$, we shall nonetheless find nontrivial effects from the unbroken $\bb{Z}_N$ factor. 

We go to low energies by imagining $v \rightarrow \infty$, in which case we should integrate out the radial mode $\rho$ of the Higgs field around the broken vacuum, and we see that the relevant coupling of the gauge and Goldstone fields dominates 
\begin{equation}\label{eqn:abelianhiggslag}
    \mathcal{L} = - v^2 \left(\partial_\mu \phi - e N A_\mu\right)^2,
\end{equation}
where corrections are of order $\mathcal{O}(v^0)$ and it is clear that in the infrared the equations of motion simply set $A = \partial \phi / N$. The gauge boson has no local degrees of freedom at low energies and is pure gauge, as is to be expected for a Higgsed vector field.

The ground state of the theory has $\langle \Phi \rangle = v$ everywhere, but our interest is in a slightly broader class of solutions. 

\subsection{Abelian Higgs Cosmic Strings} \label{sec:ahstrings}

We search for time-independent, $z$-independent solutions to the Higgs-Yang-Mills equations of motion with finite energy per unit length, which simplifies things to a $2$d problem. Of course the ground state with uniform vev satisfies this, but it is not the only one. In general the cross-sections look like Abrikosov-Nielsen-Olesen vortices \cite{Abrikosov:1956sx,Nielsen:1973cs}.

\begin{figure}[t]
    \centering
        {\includegraphics[clip, trim=0.0cm 0.0cm 0.0cm 0.0cm, width=0.5\textwidth]{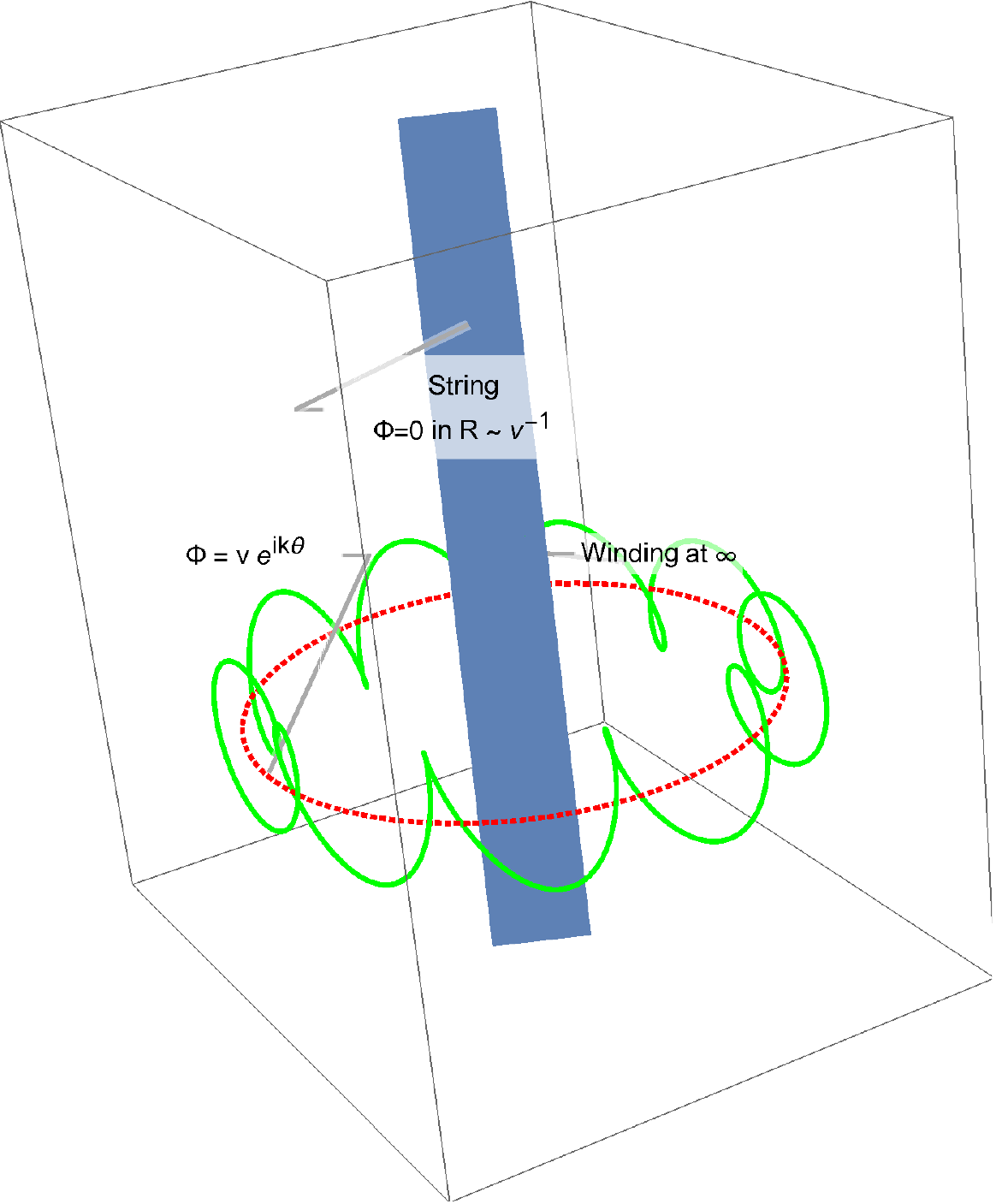}}
    \caption{Cartoon of a cosmic string solution with the winding of the Higgs phase plotted on a transverse circle.}
    \label{fig:string}
\end{figure}

With the scalar potential $V(\Phi)$ minimized for some $|\Phi|=v\neq 0$, finite energy requires toward infinity in the transverse directions
\begin{equation}
    \Phi(r,\theta) \rightarrow v e^{i\phi(\theta)}
\end{equation}
where we may view $e^{i\phi(\theta)}$ as a map from the $S^1$ at transverse infinity to the $S^1$ of the gauge group. Such maps can be topologically classified by their winding number $k \in \bb{Z}$, the canonical example of which is $e^{ik\theta}$.  Since $\Phi$ has charge $N$, the transformation $\phi \mapsto \phi + N \theta$ is a gauge symmetry and winding is only physical mod $N$. 

The equations of motion for the gauge field then dictate setting at long distances
\begin{equation} \label{eqn:stringPot}
    \vec{A}(r,\theta) \rightarrow \frac{\phi'(\theta)}{e N r} \hat{\theta}.
\end{equation}
With this we may directly relate the magnetic flux in the string to the winding number of the scalar using Stokes' theorem. 
Starting with the definition of the winding number, we have
\begin{align}
    k = \frac{e N}{2\pi} \oint_{\gamma} \vec{A} \cdot d\vec{x} &= \frac{e N}{2\pi} \int \hspace{-5pt} \int_{S} \vec{\nabla} \times \vec{A} \cdot d\vec{S} =  \frac{e N}{2\pi} \Phi_B    \\
    &\Rightarrow \Phi_B = \frac{2\pi}{e N} k, \label{eqn:magflux}
\end{align}
where we have simply used Stokes' theorem to relate the winding number to the magnetic flux $\Phi_B$ through the string. 
We see the flux carried by the string is fractionally quantized in $1/N$ of the Dirac flux quantum, $g_D = 2\pi/e$ \cite{Dirac:1931kp,Dirac:1948um}. The strings then contain only a fraction of the flux around a Gaussian sphere of a fundamental magnetic monopole.

With nonzero winding number of the scalar field, these are then cosmic string solutions. The continuity of the field demands that somewhere in the interior $\Phi=0$ and the symmetry is unbroken. While the ultraviolet form of the solutions may be found in a given theory without much difficulty, we will here only need what could be guessed by dimensional analysis: there is a core of size $\sim v^{-1}$ inside which the Higgs is pushed up to the origin and the $U(1)$ is unbroken, and the strings have tension $E/L \equiv \mu \sim v^2$. 

\subsection{Discrete Aharonov-Bohm Effect} \label{sec:aharonovbohm}

\begin{figure}[t]
    \centering
        {\includegraphics[clip, trim=0.0cm 4.0cm 1.0cm 3.0cm, width=0.5\textwidth]{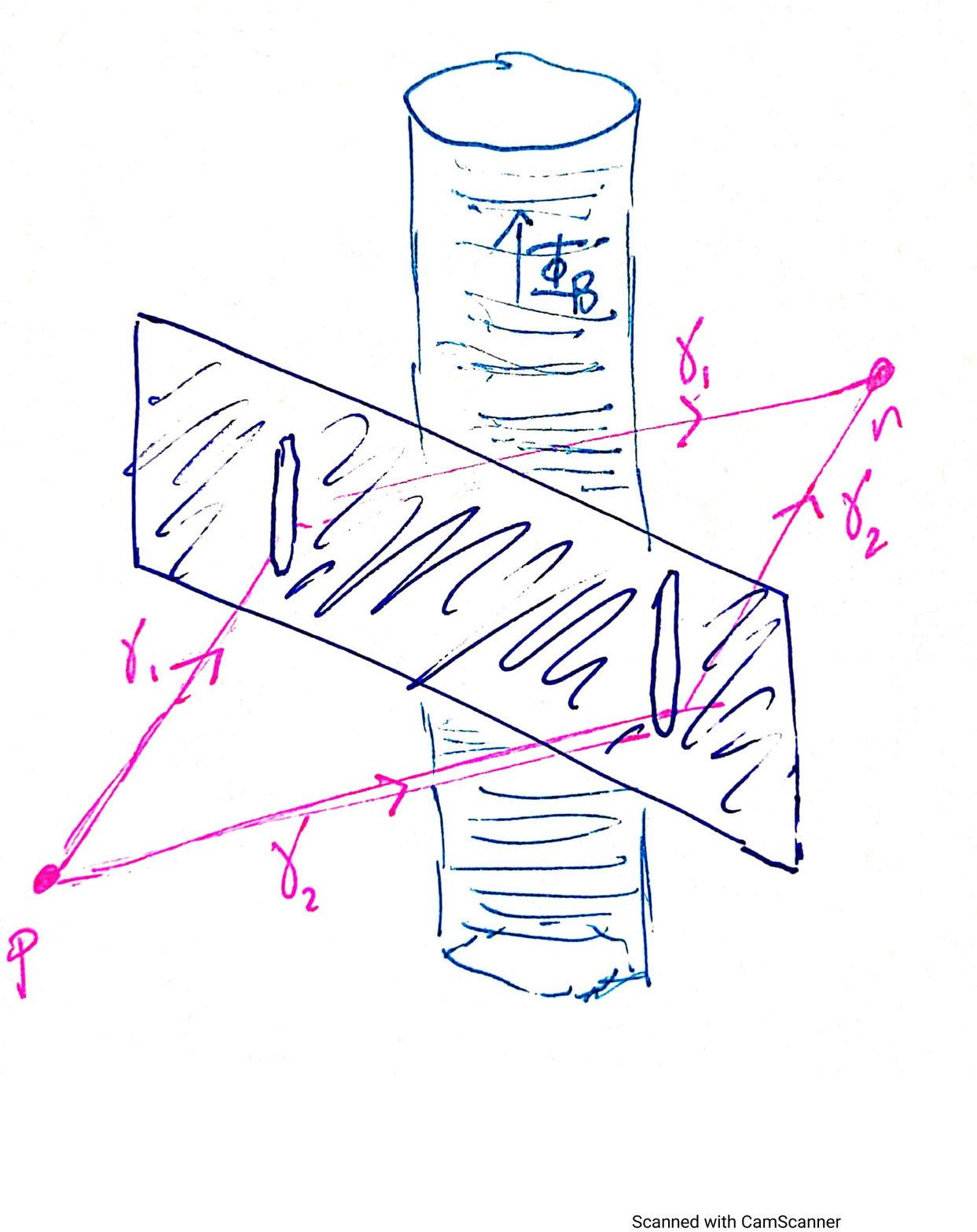}}
    \caption{Cartoon of Aharonov-Bohm effect setup with two slits on either side of a solenoid providing inequivalent paths for electrons toward an interferometer.}
    \label{fig:ABeffect}
\end{figure}

The cosmic string solutions we've described are cylindrical configurations with magnetic flux confined within a radius $R \sim v^{-1}$. This is a nearly-idealized solenoid, and indeed our long-distance solution for the vector potential in Eqn \ref{eqn:stringPot} is precisely that of the elementary Aharonov-Bohm setup \cite{Aharonov:1959fk}. As they did, we consider shooting a charged fermion from point $p$ with initial wavefunction $\psi_p$ past a solenoid to an interferometer at point $r$, and we can write the resulting wavefunction as an integral over paths $\gamma$ from $p$ to $r$
\begin{equation}
    \psi_r = \int_p^r \mathcal{D} \gamma e^{iS(\gamma)} \psi_p
\end{equation}
where $S(\gamma)$ is the action evaluated for the path $\gamma$. We diagram this in Figure \ref{fig:ABeffect}. The charged fermion has an electric current which in the nonrelativistic limit leads to an action contribution $\int d^4x A_\mu J^\mu \rightarrow q \int_\gamma A$. We assume no other terms are relevant. If the path integral has two saddle points $\gamma_1, \gamma_2$, for example due to a screen with two slits between $p$ and $r$, then at lowest order $\psi_r$ is the superposition of these two paths
\begin{equation}
    \psi_r \propto \left( e^{i q \int_{\gamma_1} A} + e^{i q \int_{\gamma_2} A} \right) \psi_p,
\end{equation}
and the probability density is then proportional to
\begin{align}
    \left| \psi_r \right|^2 &\propto \half \left(2 + e^{i q \int_{\gamma_1} A - i q \int_{\gamma_2} A} + e^{i q \int_{\gamma_2} A - i q \int_{\gamma_1} A}\right) \left| \psi_p \right|^2 \\
    &\propto \left( 1 + \text{Re} \ e^{i q \oint_{\gamma} A}\right),
\end{align}
where $\gamma$ is the loop consisting of the path $\gamma_1$ composed with the path $\gamma_2^{-1}$. We see that the two paths interfere to the extent that $\exp i q \oint_\gamma A$ differs from one---that is, due to the presence of magnetic flux in the region encircled by $\gamma$.

In other language, the existence of the magnetic flux has induced a holonomy in the gauge field $A$, $\text{Hol}_\gamma(A) \equiv \exp i \oint_\gamma A \neq 0$. This is precisely the observable of the Aharonov-Bohm effect.
\footnote{I refrain from fiber bundle language in this manuscript, but I cannot resist mentioning the explanations by Socolovsky and collaborators of the Aharonov-Bohm effect in full geometric power \cite{Socolovsky:2002a,Socolovsky:2007a}.}
A particle of electric charge $e$ transported around the solenoid then picks up a phase $\exp i \left( e \Phi_B \right)$. In the case of electromagnetism this is not quantized, since the magnetic flux through an experimentalist's solenoid is ultimately proportional to the velocity of the electrons in the wires, which is a continuous parameter. But when electromagnetism is broken, as in a superconductor, or as above from an elementary Higgs field, magnetic flux is quantized into vortices. The AB effect for the Abelian Higgs with unbroken $\bb{Z}_N$ symmetry results in a phase from Eqn \ref{eqn:magflux} of 
\begin{equation}
    \exp{i e \oint_\gamma A} = \exp{i e \frac{2\pi}{Ne}} = \exp \frac{2\pi i k}{N},
\end{equation}
and we see that for a fundamental $N=1$ Higgs the Aharonov-Bohm effect vanishes in the broken phase. The $U(1)$ charge is completely screened and the vortices have magnetic flux $\Phi_0 \equiv 2\pi/e$, which we recognize also as the flux through a sphere around a monopole with the Dirac charge quantum. 

However, when the $U(1)$ symmetry is broken by a Higgs of charge $N>1$---such that the discrete $\bb{Z}_N$ gauge symmetry remains unbroken---the magnetic flux tubes are \textit{fractionally} quantized with fundamental unit $\Phi_B = \Phi_0/N$. Then the Higgs condensate can only screen electric field lines in $N$-fold units, such that the charge $(\text{mod }N)$ is unscreened and a long-range $\bb{Z}_N$ gauge field remains \cite{Krauss:1988zc,Alford:1989ch,Preskill:1990bm,Coleman:1991ku,Banks:2010zn}. A discretized version of the Aharonov-Bohm effect must therefore persist in the infrared. And indeed, we have seen that the winding of the Higgs field provides for dynamical solitons with discrete magnetic flux---precisely a fundamental particle version of the idealized solenoid.

\begin{figure}[t]
    \centering
        {\includegraphics[clip, trim=0.0cm 0.0cm 0.0cm 0.0cm, width=1.0\textwidth]{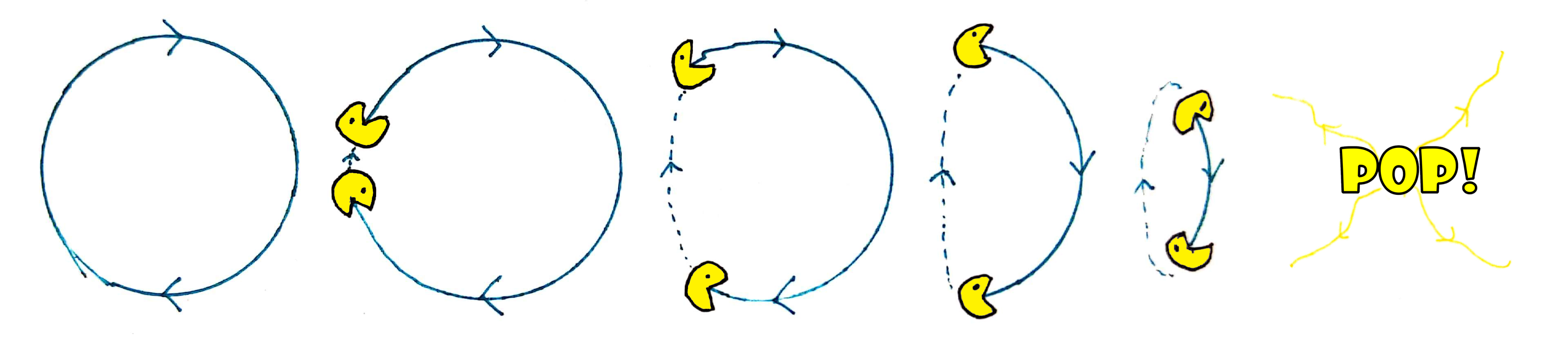}}
    \caption{The so-called `pac-man' instability of the minimal Abelian Higgs cosmic string loop to the nucleation of a monopole-antimonopole pair. With inspiration from R. Lichtenstein.}
    \label{fig:pacman}
\end{figure}

Importantly this remaining discrete gauge symmetry implies their infrared stability even in the presence of magnetic monopoles. If we consider chopping off the top half of our solenoid to have a semi-infinite solenoid running off along the negative $z$ axis, at the origin we have a source of $\Phi_B$ magnetic flux. If $\Phi_B = \Phi_0$, then in the limit when the radius vanishes $R \rightarrow 0$, the solenoid becomes the unphysical Dirac string of Dirac's $U(1)$ monopole. Then along a cosmic string one may imagine nucleating a monopole-anti-monopole pair which are pulled apart by the cosmic string tension and eat it up, as in Figure \ref{fig:pacman}. But these monopoles simply cannot remove any fractional flux $\Phi_B/\Phi_0 = k/N \ (\text{mod } 1)$.  

While the above experiment with solely two paths for the charged particle is a conceptually simple demonstration of the Aharonov-Bohm effect, the cosmologically relevant problem is simply of an incident charged particle on the solenoid. Building off earlier studies of AB scattering \cite{Aharonov:1984zza,Rohm:1985aa,deSousaGerbert:1988kpn,deSousaGerbert:1988qzd,Alford:1988sj,Wilczek:1981du}, including the original result of Aharonov and Bohm for a general solenoid \cite{Aharonov:1959fk}, Alford \& Wilczek looked at an example of cosmic strings from the breaking $SO(10) \rightarrow SU(5) \times \bb{Z}_2$ \cite{Alford:1988sj}. The generalization of their result is a differential cross-section per unit length of
\begin{equation}
    \frac{d\sigma}{d\theta} = \frac{\sin^2(\pi q \Phi_B/\Phi_0)}{2\pi p \sin^2(\theta/2)},
\end{equation}
with $p$ the incoming particle momentum, $q$ its charge, and $\Phi_B/\Phi_0$ the magnetic flux in fundamental units. 
There is then a large cross-section for elastic scattering due to the nontrivial gauge potential. The fundamental cosmic string has $\Phi_B/\Phi_0 = 1/N$, so that the Aharonov-Bohm scattering affects any particle with $q \neq 0 \ (\text{mod } N)$. 

The maximum is reached for $q = N/2$, the largest possible charge in $\bb{Z}_N$, but in our model with $N = 2 N_c \gen$ the suppression of elastic scattering from the $\sin^2$ is minor. For quarks with $q=1$, we have $\sin^2(\pi/18) \sim 3 \times 10^{-2}$, and proton or lepton scattering is only suppressed by $\sin^2(\pi/6) = 1/4$. Indeed, the cosmic strings have a large elastic scattering cross-section with all of the SM fermions. Such scattering affects neither the identity or energy of the scattered particles, but will lead to a drag force on cosmic strings moving in a dense medium. 

\subsection{$\bb{Z}_N$ Gauge Theory} \label{sec:zngauge}

In the far infrared of the $N$-Abelian Higgs when only the discrete $\bb{Z}_N$ gauge theory remains, the physics can be made more transparent by dualizing $\phi$ to a two-form gauge field $B_{\mu\nu}$, as first done in \cite{Banks:1989ag,Coleman:1991ku} and later emphasized in \cite{Banks:2010zn}. Beginning with the infrared Lagrangian of Equation \ref{eqn:abelianhiggslag}, we first rewrite it as 
\begin{equation}
    \mathcal{L} = - i J^\mu (\partial_\mu \phi - e N A_\mu) - \frac{1}{2 v^2} J^\mu J_\mu,
\end{equation}
which can be seen to produce the same Lagrangian upon setting $J^\mu$ to its equations of motion $J_\mu = -iv^2 (\partial_\mu \phi - e N A_\mu)$. If we instead integrate the first term by parts, then we can interpret $\phi$ as merely a Lagrange multiplier enforcing the constraint $\partial_\mu J^\mu = 0$.

We can trivialize this constraint on the dual current by changing variables as 
\begin{equation}
    J_\mu = - \frac{i}{2\pi} \epsilon_{\mu\nu\rho\sigma} \partial^\nu B^{\rho\sigma}
\end{equation}
where $\epsilon_{\mu\nu\rho\sigma}$ is the Levi-Civita symbol and $B$ is an antisymmetric two-tensor field. 
The two-tensor $B_{\mu\nu}$ is invariant under a gauge transformation by an exact two-form $B_{\mu\nu} \rightarrow B_{\mu\nu} + \partial_{\left[\mu\right.} \lambda^{(1)}_{\left.\nu\right]}$ parametrized by a one-form $\lambda^{(1)}$ in a higher-dimensional analogue of the one-form gauge symmetry of electromagnetism $A_\mu \rightarrow A_\mu + \partial_\mu \lambda^{(0)}$.

When we then rewrite the Lagrangian in terms of this two-form and look at low energies we find 
\begin{align} \label{eqn:BFlag}
    \mathcal{L} &= \frac{e N}{2\pi} A^\mu \partial^\nu B^{\rho\sigma} \epsilon_{\mu\nu\rho\sigma} \\
    &= - \frac{N}{2\pi} B \wedge F, \nonumber
\end{align}
the so-called `BF theory' first introduced by Horowitz \cite{Horowitz:1989ng,Blau:1989bq,Blau:1989dh,Wu:1990ci}, where in the second line we've cleaned up the indices by instead writing a wedge product of two-forms (and also put the coupling back into $A$). In the IR the kinetic terms for the $A,B$ fields disappear, meaning that there are no local degrees of freedom, as expected. 

Nonetheless, the theory is not entirely trivial: The BF theory is a topological theory of an Abelian one-form gauge field $A$ and an Abelian two-form gauge field $B$. With just our two gauge fields, we have as observables the Wilson lines $\exp i n_A \int_{\Sigma_1} A$ and Wilson surfaces $\exp i n_B \int_{\Sigma_2} B$, where $\Sigma_d$ is a $d$-dimensional surface and $n_A, n_B$ are integers. These capture the effects of the gauge field on the phase of the wavefunction for a probe particle which has no other dynamics. As effective field theorists, we may think of these as the infrared remnants of massive charge-$n_A$ particles with worldline $\Sigma_1$ or massive strings of winding $n_B$ with worldsheet $\Sigma_2$. Indeed, the one-dimensional objects charged under the two-form gauge symmetry are none other than the cosmic strings we studied above. Our absorption of the charge $e$ into the gauge field $A$ puts the Wilson operators on equal footing in terms of solely the integers $n_A, n_B$ appearing. Alternatively we could have rescaled $B$ to make explicit its coupling, which is $v^{-1}$.

\begin{figure}[t]
    \centering
        {\includegraphics[clip, trim=0.0cm 0.0cm 0.0cm 0.0cm, width=0.5\textwidth]{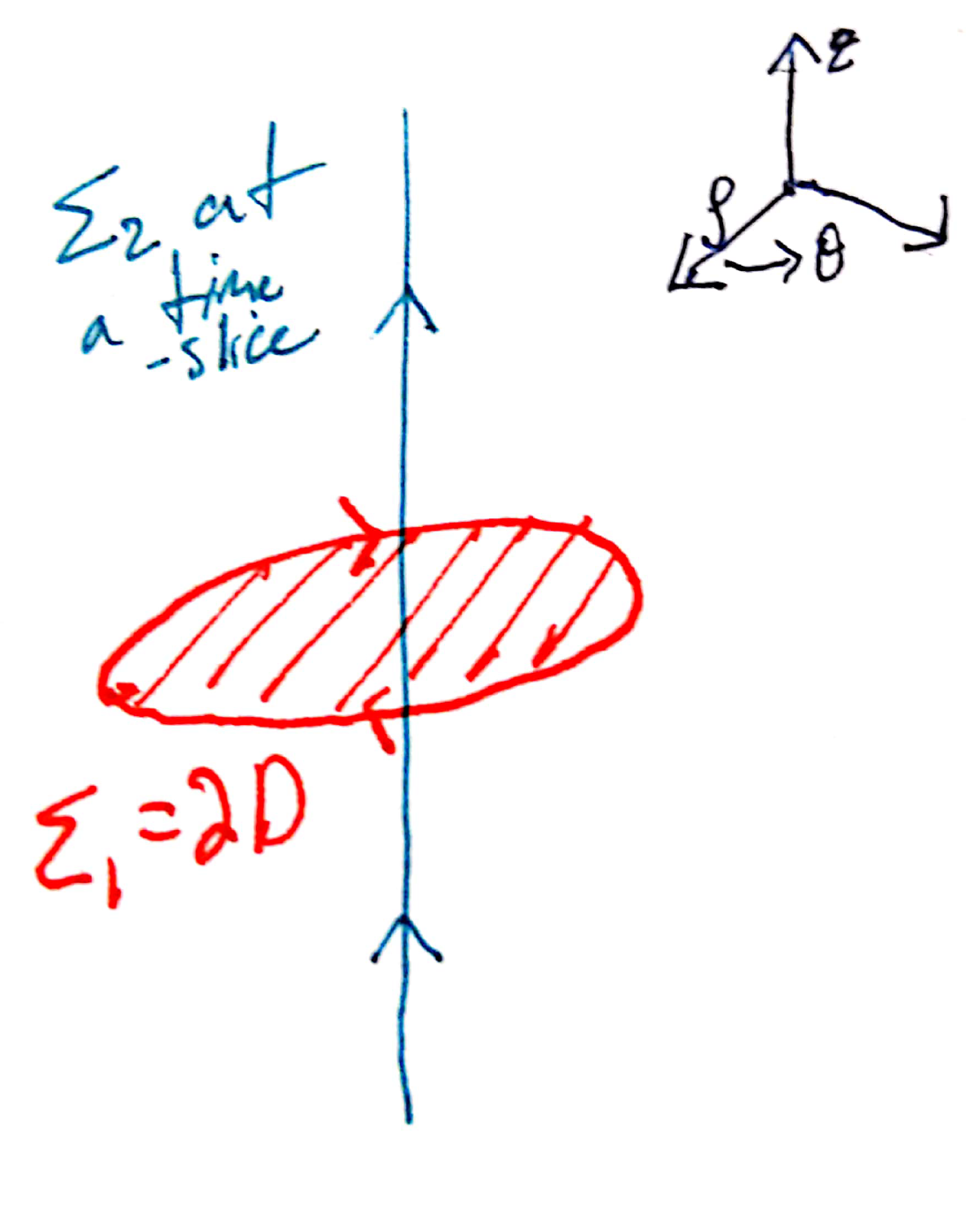}}
    \caption{Example configuration for a two-observable correlator in the BF theory with $\Sigma_1$ spacelike and $\Sigma_2$ at $x=y=0$.}
    \label{fig:BFlink}
\end{figure}

Furthermore, the topological interaction gives precisely the appropriate Aharonov-Bohm phases as these wind around each other, and ensures that only $n_A n_B \ (\text{mod } N)$ is observable. This is the signature of the $\bb{Z}_N$ gauge symmetry at low energies. One can calculate the expectation value of a Wilson loop and a Wilson surface to find
\begin{equation} \label{eqn:bfcorr}
    \left\langle e^{i n_A \int_{\Sigma_1} A} \  e^{i n_B \int_{\Sigma_2} B} \right \rangle = \exp{\left[2\pi i\frac{n_A n_B}{N} \text{link}(\Sigma_1,\Sigma_2)\right]},
\end{equation}
where $\text{link}(\Sigma_1,\Sigma_2)$ is the linking number of these two manifolds and in some sense this can be used to define generalized linking numbers quantum field theoretically \cite{Oda:1989tq,Horowitz:1989km}. This can be understood by recognizing that BF theories are naturally theories of de Rham cohomology groups of fiber bundles, but we can convince ourselves of this in a more elementary fashion with inspiration from our above discussion of the AB effect.

Our strategy is to push the Wilson surface operator into the action, and reinterpret this as the expectation value of the Wilson loop in a background of a probe string. We let $X^a$ be coordinates on $\Sigma_2$, and $x^\mu = \xi^\mu(X^a)$ the position of the surface in spacetime, and set $n_A=n_B=1$ for cleanliness. We have
\begin{equation}
    \left\langle e^{ i \int_{\Sigma_1} A} e^{ i \int_{\Sigma_2} B} \right\rangle = \int \mathcal{D}A \mathcal{D} B e^{ i \int_{\Sigma_1} A} e^{i\left(S_{BF}+ \int_{\Sigma_2} B\right)}, 
\end{equation}
where the action has been modified to
\begin{equation}
    S_{BF+B} = \int d^dx \ \left(- \epsilon_{\mu\nu\rho\sigma}  \frac{N}{2\pi} \partial^\mu A^\nu B^{\rho \sigma}
    +  B_{\mu\nu} \delta^{(2)}\left(x - \xi(X)\right) \frac{\partial \xi^\mu}{\partial X^a} \frac{\partial \xi^\nu}{\partial X^b} \epsilon^{ab},
    \right) 
\end{equation}
and we can reinterpret this as adding to our Lagrangian an operator to create an extended object, which the delta function localizes to a submanifold \cite{Brennan:2021a}. We can make the simple gauge choice $\xi^\mu(X) = \delta^\mu_a X^a$ to make the Jacobian trivial.
Now by examining the equation of motion from varying $B$, we see that this term induces a fractional flux for the gauge field $A$, 
\begin{equation} 
    \frac{\delta \mathcal{L}}{\delta B^{\sigma\rho}} =  - \epsilon_{\mu\nu\rho\sigma} \frac{N}{2\pi} \partial^\mu A^{\nu}  + \delta^{(2)}\left(x - X\right) \delta_\rho^a \delta_\sigma^b \epsilon_{ab} = 0.
\end{equation}
To make the physical picture clearer, let's specialize to the case of $\Sigma_2$ the worldsheet of a static cosmic string at $X^a=(t,z)$, and $\Sigma_1$ a spatial loop which encircles the $z$-axis, as in the Aharonov-Bohm setup. Then the equation of motion for $\rho = t, \sigma = z$ gives us
\begin{equation}
    \vec{\nabla} \times \vec{A} = \frac{2\pi}{N} \delta\left(x\right) \delta\left(y\right) \hat{z},
 \end{equation}
and this is nothing other than an induced fractional holonomy for the gauge field around the string. If we integrate this over a spatial surface $D$ with $\partial D = \Sigma_1$, then it is clear that if $\Sigma_1$ loops around the spatial $S_1$ of $\Sigma_2$ we have 
\begin{equation}
    \oint_{\Sigma_1} A = \int_D \nabla \times A =   \frac{2\pi}{N}.
\end{equation}
Then we can use this modified equation of motion for $A$ in evaluating the expectation value and find
\begin{equation}
    \left\langle e^{ i \int_{\Sigma_1} A} \right\rangle_{BF+B} = e^{2\pi i/N},
\end{equation}
precisely the holonomy we saw in Appendix \ref{sec:aharonovbohm} appearing for a unit-charged particle transported around a fundamental cosmic string in the non-minimal Abelian Higgs. The general result of Equation \ref{eqn:bfcorr} should then be thought of as the relativistic generalization of the discrete Aharonov-Bohm effect to arbitrary line and surface operators.

Indeed, the cosmic strings of the Abelian Higgs theory source the two-form gauge field $B_{\mu\nu}$, and the Wilson surfaces are their remnant in the IR $\bb{Z}_N$ gauge theory as non-dynamical sources. Just as in the Abelian Higgs description, the strings induce a holonomy $e^{2 \pi i \Phi_B/\Phi_0} \in \bb{Z}_N$ for the gauge field $A$. We will return to the BF theory in Appendix \ref{sec:catalysis}, but its full application to our scenario will be further studied elsewhere. In the meantime, reviews of discrete gauge theories include \cite{deWildPropitius:1995hk,Alford:1991vr,Birmingham:1991ty}.

\section{Cosmic String Dynamics} \label{sec:cosmicstrings}

In this section we review some qualitative features of the formation and evolution of cosmic strings. Ultimately our interest will be in an early-time and small-scale regime where the details and not yet known, so we will resort simply to parametrizing the string number density, as above. But understanding the features below is still a useful guide to their behavior.

\subsection{Cosmic String Evolution} \label{sec:evolution}


In the cosmological case of a dynamical phase transition in the early universe, the production of cosmic strings is \textit{guaranteed} by the finite size of the cosmic horizon. The field $\Phi$ condenses with independent phases in different Hubble patches, so line defects necessarily form. The string formation may be thought of as a random walk, which in three spatial dimensions need not return to its starting point. So an order one number of super-Hubble length cosmic strings per Hubble volume is expected to form by simple dimensional analysis.  This was first studied by Kibble \cite{Kibble:1976sj,Kibble:1980mv} and the formation of such topological defect solutions is sometimes known as the `Kibble mechanism'. Numerical simulations thereafter \cite{Vachaspati:1984dz} confirmed that an order-one fraction of the energy density in strings during formation is in infinite, horizon-crossing strings. 

Resultingly, the exotic solutions to the Abelian Higgs equations of motion that we constructed in Sec \ref{sec:ahstrings} are not merely of formal interest, but are necessarily realized in the early universe. In the following section we will discuss the evolution of these objects after the phase transition in which they are formed. The strings are not charged under any unbroken continuous gauge symmetries, so there are no massless gauge bosons mediating long-range forces. As a result, the long-range dynamics of an isolated string are purely gravitational and controlled by the classical Nambu-Goto action of a gravitating string \cite{Nambu:1970a,Goto:1971ce,Forster:1974ga}.

We note that there is lots of physics in the details of the Kibble mechanism and the spectrum of defects formed, and this connects closely to condensed matter realizations thereof. In particular, the correlation length of Higgs fluctuations may in fact be quite small compared to the horizon length $\xi \ll d_h$, resulting in a far greater initial number density of defects \cite{Guth:1982pn,Zurek:1985qw,Murayama:2009nj}. Our scenario will need to rely on this early-time overabundance of strings, but this is not yet well-understood. 

The late-time behavior is still non-trivial and quite interesting. There is an infamous result that cosmic strings reach an attractor solution where the string network has a scaling symmetry---the distribution of string loop lengths depends solely on the length as a fraction of the Hubble radius. This was first argued for on energetic grounds by Turok and collaborators \cite{Albrecht:1984xv,Turok:1984db}, with the only other possibility being overclosing the universe \cite{Kibble:1984hp}. Simulations by Bennett \& Bouchet provided numerical evidence for the scaling regime near the horizon scale shortly thereafter  \cite{Bennett:1987vf,Bennett:1989ak,Bennett:1989yp,Sakellariadou:1990nd,Allen:1990tv}. Analytic arguments for the scaling solution were later given by Kibble \& Copeland \cite{Martins:1996jp,Kibble:1990ym,Copeland:1991kz,Austin:1993rg}, but confusion reigned about the situation on small scales, with some simulations finding copious production of small loops \cite{Vincent:1996rb} and uncertainty about the role of `kinks' and other structure on the strings \cite{Kibble:1990ym,Siemens:2001dx,Siemens:2002dj}. Increasingly powerful simulations have further confirmed the scaling regime \cite{Moore:2001px,Ringeval:2005kr,Martins:2005es,Vanchurin:2005pa,Olum:2006ix,Blanco-Pillado:2011egf,Blanco-Pillado:2013qja,Hindmarsh:2008dw,Auclair:2019zoz}. A microphysical explanation in terms of the classical string theory was finally provided by Polchinski \& Rocha \cite{Polchinski:2006ee,Polchinski:2007rg,Rocha:2007ni,Dubath:2007mf,Lorenz:2010sm}, with \cite{Polchinski:2007qc} giving a nice overview of the difficulties of this problem.

What this means is that the late-time behavior is controlled by strings chopping each other up into loops and ensmallening by emitting gravitational radiation. We will discuss these effects in the next two sections. 

However, before the late-time attractor dynamics can take over, there is a period when friction from elastic Aharonov-Bohm interactions with the SM plasma dominates. That cosmic strings dominantly interact with matter through the Aharonov-Bohm effect was first appreciated only in 1989 by Alford \& Wilczek \cite{Alford:1988sj}. As we saw in Sec \ref{sec:aharonovbohm}, the string elastically scatters charges at an enormous rate $\sigma/\ell \sim 1/p$, with $p$ the momentum.  This scattering from the gauge potential is far more important than the friction due to gravitational effects which had previously been considered \cite{Everett:1981nj,Vilenkin:1984ib,Silva:2021www}.

Vilenkin then showed \cite{Vilenkin:1991zk} that in the rest frame of a straight section of string moving at $\vec{v}$ with respect to the background gas of particles with charge $q$ at temperature $T$, the frictional force is 
\begin{equation}
    \vec{F} \simeq -T^3 \frac{\vec{v}}{\sqrt{1-v^2}} \sin^2(\pi \frac{q k}{N}).
\end{equation} 
where I've gotten rid of some numerical factors. The effect on the string is to damp out motion on a timescale $\propto \mu/T^3$. This frictional force turns out to severely damp the dynamics until $T \lesssim (G \mu) \mpl$, when the strings can begin free-streaming. See also \cite{Garriga:1993gj,Martins:1995tg}.

While a single, long, straight string is easy enough, an understanding of the dynamics when they free-stream quickly becomes very complicated. There are an enormous range of scales involved in the problem---from the Hubble scale, to the interstring distance, to the curvature length along the string, down to the fundamental transverse size of the string. For fundamental cosmic strings, the interactions are controlled by a string coupling which may be small. However, our strings are semi-classical and have no free limit. As a result, the strings strongly self-interact. The Polyakov dynamics \cite{Polyakov:1981rd} break down entirely when strings intersect each other, or themselves, and the results of these interactions are crucial for understanding the network dynamics. 

For completeness, I briefly note here the interesting program of understanding the cosmic string behavior using the worldsheet theory derived from the Abelian Higgs theory (e.g. \cite{Gregory:1988qv,Akhmedov:1995mw,Gregory:1990pm,Silveira:1993am,Orland:1994qt,Arodz:1995dg,Baker:1999xn,Baker:2000ci,Aharony:2013ipa,Szabo:1998ej,Bergeron:1994ym,Zee:1994qw,Sato:1994vz}).

\subsection{Intercommutation}\label{sec:intercom}

\begin{figure}[h]
    \centering
        {\includegraphics[clip, trim=0.0cm 7.0cm 0.0cm 7.0cm, width=\textwidth]{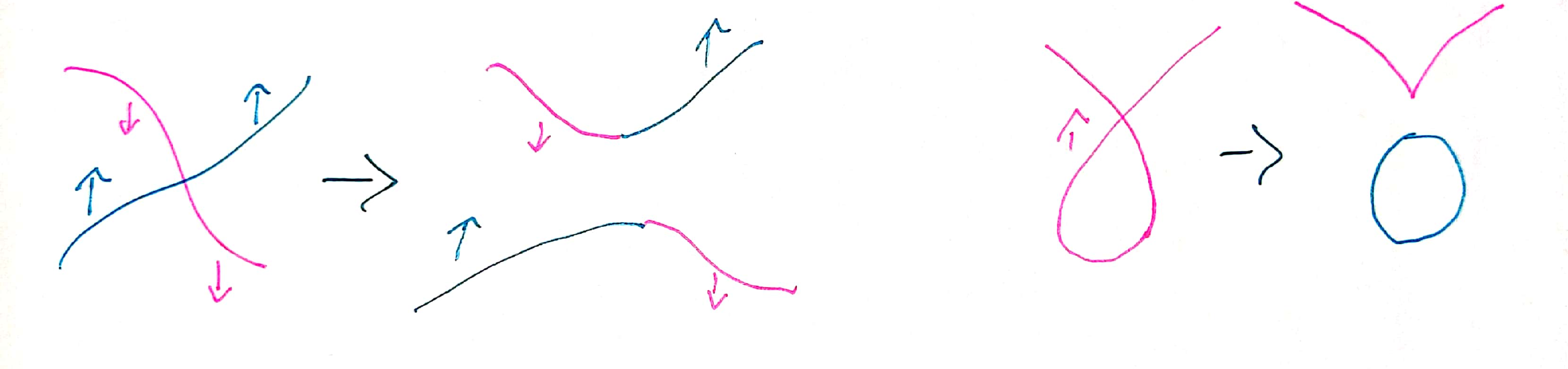}}
    \caption{Cartoon of intercommutation dynamics whereby colliding string exchange partners,and self-intersecting strings chop off loops.}
    \label{fig:intercom}
\end{figure}

There is a counter-intuitive effect in string scattering which turns out to be crucial to our phenomenology: colliding Abelian Higgs cosmic strings always `intercommute' (or `reconnect' or `trade ends' or `exchange partners'), as diagrammed in Figure \ref{fig:intercom}. 
This effect has been seen numerically to apply quite broadly, but as far as I am aware there is analytic understanding only in the case of supersymmetric defects, so we discuss that case to gain some qualitative conceptual insight.

That understanding begins with Ruback \cite{Ruback:1988ba}, who found in $2+1$d that identical vortices undergo right-angle scattering by following the moduli space approximation introduced by Manton \cite{Manton:1981mp} for monopoles and eventually rigorously justified by Stuart \cite{Stuart:1994tc,Stuart:1994yc}, see also \cite{Atiyah:1985dv,Gibbons:1986df,Rosenzweig:1990ea,Samols:1991ne,Abdelwahid:1993fg,MacKenzie:1995np,Gauntlett:2000ks}. Topological defects at low energy follow geodesics on the moduli space, with a metric derived from the kinetic terms in the Lagrangian. 

In the case of two identical vortices colliding at low energies, the only relevant coordinate on the two-vortex moduli space is the relative position $\vec{r}$ pointing from the zero in the Higgs field of one to the other. So the moduli space in the winding number $k=2$ sector looks locally like the plane $\bb{R}^2$. But despite being semi-classical objects, two vortices may be identical and so indistinguishable, so $-\vec{r}$ describes the same configuration. Then the moduli space has the geometry of a cone with deficit angle $\pi$---at least at large separations. But it's more difficult to reason about the behavior near the tip, since at small distances $|r| \lesssim v^{-1}$ the vortices lose their separate identities and our above argument fails.

\begin{figure}[t]
    \centering
        {\includegraphics[clip, trim=0.0cm 0.0cm 1.0cm 0.0cm, width=1.0\textwidth]{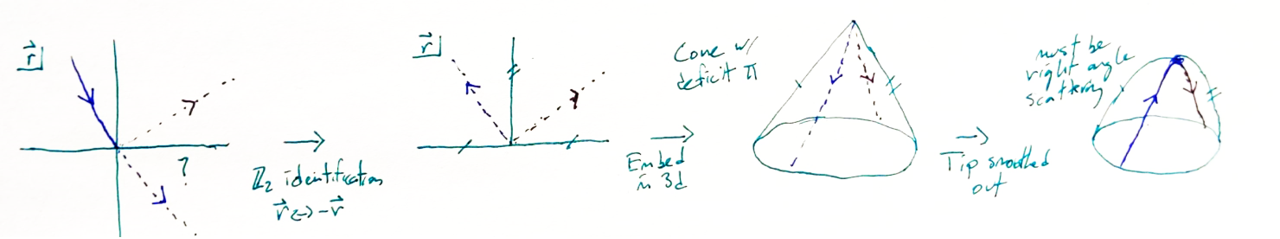}}
    \caption{Cartoon of motion in the two-vortex moduli space. Since the vortices are indistinguishable, the space is a cone, though the tip is smoothed out. The radial geodesic then clearly passes straight over the cone, which in physical space corresponds to right-angle scattering.}
    \label{fig:modSpaceCone}
\end{figure}

In fact the moduli space is known to be smooth  \cite{Taubes:1979tm} so that the cone \textit{must} be rounded off, and indeed in solvable UV-completions one can explicitly see the tip of the cone smoothed out \cite{Hanany:2005bc,Hashimoto:2005hi}. Then the intuitive picture is clearer, and we've diagrammed the situation in Figure \ref{fig:modSpaceCone}. An ingoing radial geodesic goes straight over the cone, and this corresponds in physical space to a scattering angle of $\pi/2$, whereas forward scattering would correspond to a geodesic `bouncing back' at the same angle after hitting $\rho=0$. We try to diagram this logic in Figure \ref{fig:modSpaceCone}.

\begin{figure}[t]
    \centering
        {\includegraphics[clip, trim=0.0cm 0.0cm 0.0cm 0.0cm, width=0.9\textwidth]{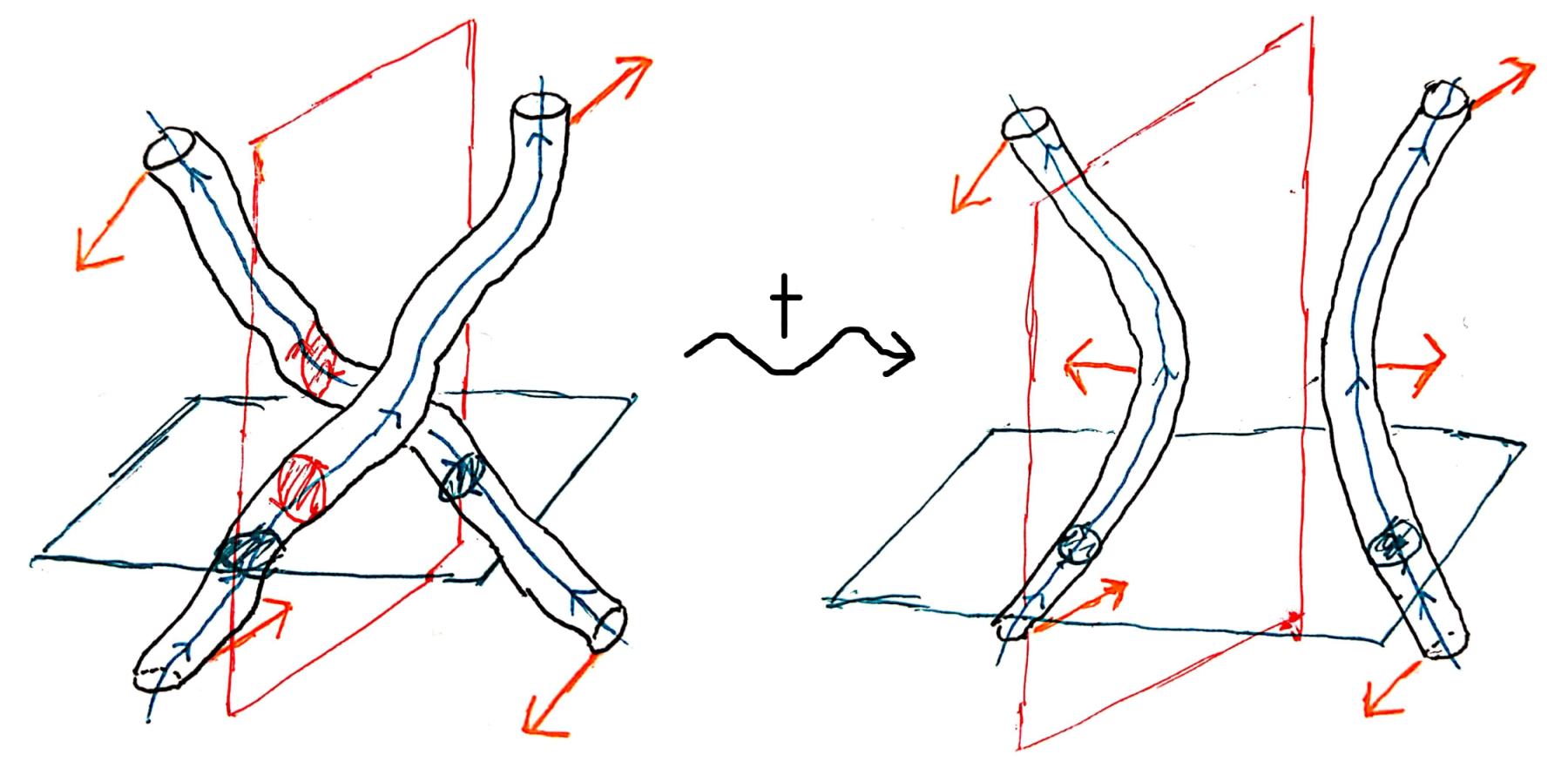}}
    \caption{Cartoon of an interaction between cosmic strings, zoomed in near the collision point. Orange arrows give the string velocities, which are initially into/out of the page. The configuration can be projected onto the red plane through which the string fluxes pass in different directions, such that the motion of the zeros of the Higgs field on this plane looks like  vortex-antivortex annihilation. On the other hand, projecting onto the green plane the motion is of right-angle vortex-vortex scattering. Consistency with these 2+1d dynamics implies that after the collision the strings have exchanged ends---they must intercommute. Heavily inspired by Figure 1 of \cite{Hashimoto:2005hi}.}
    \label{fig:stringCollision}
\end{figure}

The $2+1$d right-angle scattering can be uplifted to cosmic string intercommutation by projecting the motion of the $3+1$d strings down to the motion of the cores in various planes \cite{Shellard:1988ki,Hashimoto:2005hi,Tong:2005un}, as in Figure \ref{fig:stringCollision}. The fluxes pass through some such planes in the same direction, in which case the motion on the plane is of vortices undergoing right-angle scattering, but the fluxes pass oppositely through other planes on which the interaction is of vortex-antivortex annihilations. Intercommutation necessarily follows from the consistency of these lower-dimensional interactions.

Numerical simulations of the underlying field theory dynamics have also been used to investigate the string dynamics during collisions \cite{Shellard:1987bv,Shellard:1988zx,Matzner:1988qqj,Moriarty:1988em,Myers:1991yh}, and indeed are in agreement with the intercommutation probability being near unity. The string network simulations we will match to use the Nambu-Goto action to evolve their numerical strings and then merely implement intercommutation by hand whenever a collision is detected. Given the enormous range of scales in the problem it seems untenable to resolve the microphysical interaction, though we note recent work on axion strings using adaptive mesh refinement methods which may help also in this case \cite{Buschmann:2021sdq}.


This intercommutation is crucial to prevent the string network from dominating the energy budget of the universe \cite{Albrecht:1984xv,Turok:1984db}. Such interactions chop strings into smaller pieces, and lead to a distribution of lengths of string loops being created at any time, as we will see in Sec \ref{sec:simulations}. The many loops chopped off in such processes allow the network to dissipate energy by emitting gravitational radiation as they oscillate. 

We note also that each intercommutation leads to the formation of `kinks'---small-scale structure originating at the interaction point and then travelling in left- and right-moving modes (see e.g. \cite{Blanco-Pillado:2015ana,Kibble:1990ym,Copeland:2009dk,Kibble:1990ym,Siemens:2001dx,Siemens:2002dj,Rosenzweig:1990ea}). We defer further understanding of the effects of this small-scale structure to future work---if anything, it seems sensible to expect this to result in a larger number of small loops produced by the network (e.g. \cite{Dubath:2007mf}), so that ignoring this is in some sense conservative. 

\subsection{Diminution} \label{sec:shrink}

Energy loss by string loop radiation is then an important effect for understanding the network dynamics \cite{Vilenkin:1981bx,Vilenkin:1981kz}. In the string rest frame the power emitted into gravitational waves from a string loop of length $\ell$ oscillating at a frequency $\omega$ is given by the `quadrupole formula' and estimated as \cite{Vilenkin:2000jqa}
\begin{equation}\label{eqn:gravrad}
    \frac{d M}{dt} \sim G \left( \frac{d^3 Q}{dt^3} \right)^2 \sim G M^2 \ell^4 \omega^6 \sim G \mu^2
\end{equation}
where $Q \sim M \ell^2$ is the approximate quadrupole moment of the string loop and $\omega \sim \ell^{-1}$ is the characteristic frequency. This results in an energy loss which is independent of the overall size of the string, and simulations have found $\dot{M} = \Gamma G \mu^2$ with $\Gamma$ a numerical coefficient of order $\Gamma \sim 50$. We note also that strings may well radiate more than just gravitons \cite{Vilenkin:1986zz,Cui:2008bd,Long:2014lxa,Long:2014mxa,Hindmarsh:2017qff,Long:2019lwl,Auclair:2019jip,Vachaspati:2009jx,Hyde:2013fia,Lunardini:2012ct,Vachaspati:2009kq,Damour:2001bk}, and this fact may be used for interesting phenomenological ends (see e.g. \cite{Bhattacharjee:1982zj,Brandenberger:1991dr,Brandenberger:1992ys,Davis:1992fm,Ma:1992ky,Lew:1993bb,Brandenberger:1994mq,Mohazzab:1994xj,Sato:1995ea,Davis:1996zg,Sahu:2004ir,Sahu:2005vu}).

On large distances the cosmic strings are expected to behave as Nambu-Goto strings, and to have universal gravitational effects which offer the best observational prospects for general cosmic strings. The gravitational effects are dominated by the long strings, which carry an order one fraction of the energy density, so our understanding of this limit has seen intense study. But our interest is in the total number density, which will be dominated by small loops. And so the behavior which has yet been understood precisely does not suffice for our interests, \textit{malheureusement}. 
As to smaller scales, 
Eqn \ref{eqn:gravrad} motivates the definition of a dimensionless `gravitational cutoff' or `gravitational backreaction scale' $\xi \sim G\mu$, such that at time $t$, a loop of length $\ell \sim G\mu t$ would disappear entirely in a Hubble time. The behavior of such small loops remains the subject or debate. 

The evolution at the smallest scales further depends on the stability properties of the strings. While in the low-energy theory the cosmic strings are topologically stable as a result of the remnant $\bb{Z}_N$ symmetry discussed above, this stability is challenged when the radial mode can jump over the barrier at the origin of field space or tunnel through it. Thus once a string loop has shrunk in size to near the symmetry-breaking scale $\ell \sim v^{-1}$, the remnant discrete gauge symmetry does not prevent it from further shrinking to nonexistence.

However, these loops may be stabilized by the presence of superconducting currents \cite{Witten:1984eb, Davis:1988ip,Davis:1988jq} if there are appropriate degrees of freedom living on their cores. These `vortons' are then stabilized against further decay after reaching a small size where the repulsive effect of angular momentum prevents further dissipation \cite{Carter:1990sm}. Interestingly, the stability properties may be affected by degrees of freedom which are heavy in vacuum, but have massless excitations on the string worldsheet. Indeed, depending on the UV field content some relic vortons are inevitable \cite{Auclair:2020wse}, which places an upper limit on the string tension $G\mu$ so the vortons do not overclose the universe \cite{Peter:2013jj}. In the case $G\mu \lesssim 10^{-15}$, stable vortons may comprise some part of the dark matter density.

Our benchmark scenario is of strings which are ultimately unstable, but below we will also calculate the number density for the case in which strings are stable. Of course a realistic model may lie somewhere between these, with a fraction of strings ending up as vortons and others decaying.

\subsection{Scaling from Simulations} \label{sec:simulations}

As we discussed in Appendix \ref{sec:evolution}, we know now with confidence that the cosmic string network has a scaling solution at distances near the horizon length. The interplay of long strings entering the horizon, intersections of strings chopping off loops, and gravitational radiation emission leads to an attractor solution with self-similar dynamics. As this is the only regime in which the evolution is under control, we will make the simplifying assumption that we enter it instantaneously at some time $t_\star$ when the scale factor is $a_\star$.

But it is important to note that many uncertainties remain. 
We will assume the scaling regime extends down to the gravitational backreaction scale, which is further than numerical simulations have been able to test. We note also that some simulations have found a lengthy `transient' regime before scaling is reached, so it is possible that $t_\star$ may be long after the phase transition. During this approach to scaling, it has been found that small loops are overproduced (e.g. \cite{Olum:2006ix}), but as far as I am aware this transient regime has not been well-characterized.

We will follow the formalism of \cite{Blanco-Pillado:2013qja} and describe the distribution of strings using the scaling length variable $\alpha \equiv m/(d_h \mu)$ with $m$ the mass of the string and $\mu$ the string tension. The scaling regime implies that the loop production as a function of cosmic time and loop mass depends only on the ratio $\alpha \propto m/t$. In the following section we will sum up the total number density of loops in the scaling regime. 

The behavior can be qualitatively understood quite simply: In scaling, the network produces a constant number of loops per Hubble spacetime volume. In a cosmology with equation of state $w > -1/3$ (e.g. radiation or matter domination), the Hubble size grows more quickly than does the physical volume. Then the same loop production is spread over a larger physical volume at later times, so it is the early production of loops that matters most. 

We use the formalism and numerical results of \cite{Blanco-Pillado:2013qja} as a benchmark. We define a momentum-integrated form on the string loop phase space as $f(t,m)=\int f(t,m,p) \text{d}p$, and then $f(t,m) \Delta t \Delta m$ gives a comoving number density of loops produced in time $t \dots t + \Delta t$ with mass between $m\dots m + \Delta m$. We wish to sum up the production of loops over time using the production function found numerically.

The results of simulations in the scaling regime are characterized in terms of a scaling production function $f(\alpha) \Delta \alpha$. At a fixed time $t$, this may be interpreted as the number of loops produced in a Hubble volume $d^3_h$ over a Hubble time $d_h$ with mass $m = d_h \mu \alpha \dots d_h \mu(\alpha + \Delta \alpha)$. We can then relate the two functions as
\begin{equation}
   \frac{f(\alpha) \Delta \alpha}{d^4_h}  = \frac{f(t,m=d_h \mu \alpha)\Delta m}{a^3} 
\end{equation}
and in the limit $\Delta m/\Delta \alpha \rightarrow \left.\frac{\partial m}{\partial \alpha}\right|_{t} = d_h \mu$ we have 
\begin{equation} \label{eqn:fixedtime}
    \left.f(\alpha)\right|_{t} = \frac{\mu d^5_h}{a^3} f(t,m=d_h \mu \alpha).
\end{equation}
\noindent Simulations have found power-law scaling behavior of the sort \begin{equation}
    f(\alpha) \text{d} \alpha \sim \frac{\text{d} \alpha }{\alpha^p} \label{eqn:prodsim}
\end{equation}
with $p > 1$, so the production is peaked toward small loops. Though simulations are limited by their resolution to explore only a few decades below the Hubble scale, we will extrapolate this production down to the gravitational backreaction scale $\xi \sim G \mu$, taking $m_{\text{min}}(t,t) = \xi \mu d_h$. 

The coefficient is by definition independent of $\alpha$, which is the only relevant parameter by the scaling symmetry, so a coefficient of order unity is indeed borne out by simulations. As for the exponent $p$, we will use $p=1.7$ as a benchmark, which is the value determined during the matter-dominated era by \cite{Blanco-Pillado:2013qja}.  As our purpose is just to evince the parametrics and show the number density of string loops may be large enough, we will not delve into the full body of work determining the precise form of scaling, work on which may be found in e.g. 
\cite{Allen:1990tv,Bennett:1985qt,Bennett:1986zn,Bennett:1987vf,Bennett:1989ak,Bennett:1989yp,Albrecht:1989mk,Hindmarsh:2008dw,Polchinski:2006ee,Polchinski:2007rg,Lorenz:2010sm,Auclair:2019zoz,Vanchurin:2005pa,Ringeval:2005kr,Martins:2005es,Blanco-Pillado:2011egf,Blanco-Pillado:2013qja,Martins:1996jp,Martins:1995tg,Vincent:1996rb,Copeland:1991kz,Blanco-Pillado:2019tbi,Rocha:2007ni,Almeida:2021ihc}.

\paragraph{Small Loop Enhancement} 

If we have a loop production function in terms of the comoving number density, then we can find the total number of loops simply by adding up the loops produced at each time. We must translate the constant production per Hubble spacetime volume to the comoving volume.
\begin{align}
    n(t) &= \int^t_{t_\star} dt' \int_{m_\text{min}(t',t)} dm' f(t',m') \\
    &= \int^t_{t_\star} dt' \int_{\alpha_\text{min}(t',t)} \frac{a'^3}{d'^5_h \mu} f(\alpha') d\alpha' d'_h \mu \\
    &= \int^t_{t_\star} \frac{dt'}{d'_h} \frac{a'^3}{d'^3_h} \int_{\alpha_\text{min}(t',t)}  f(\alpha') d\alpha'
\end{align}
If loops are stable, we can simply sum over all loops produced at each time, taking $\alpha_{\text{min}}(t',t) \equiv \xi$, the gravitational cutoff scale. But if loops are unstable, we must only sum over those which will not have disappeared by $t$. Then we should take 
\begin{align}
    m_{\text{min}}(t',t) &= \xi \mu \left(d_h +t-t'\right) \\
    \alpha_{\text{min}}(t',t) &= m_{\text{min}}(t',t)/(\mu d'_h)
\end{align}
for which a loop produced with scaling mass $\alpha_{\text{min}}$ at time $t'$ has shrunk to the cutoff scale $\xi$ at time $t$. Then with a power-law loop production function we have
\begin{equation}
    n(t) = \int^t_{t_\star} \frac{dt'}{d'_h} \frac{a'^3}{d'^3_h} \xi^{1-p}
\end{equation}
in the stable case, and 
\begin{equation}
    n(t) = \int^t_{t_\star} \frac{dt'}{d'_h} \frac{a'^3}{d'^3_h} \xi^{1-p} \left[ \frac{\left(d_h +t-t'\right)}{d'_h} \right]^{1-p}
\end{equation}
when loops are ultimately unstable. When solely one sort of background evolution is relevant, we have $d'_h = t'/(1-\nu)$ and $a' = a_\star (t/t_\star)^\nu$. During radiation domination, $\nu = 1/2$, and during matter domination, $\nu = 2/3$.

In the stable case the integral may be done immediately, to find
\begin{align}\label{eqn:stabnum}
    n(t) \simeq \xi^{1-p} \frac{a_\star^3 }{t_\star^{3}}
\end{align}
where we have left off some order-one factors involving $\nu$ for readability, and ignored a term which vanishes as $t_\star/t \rightarrow 0$.
We see that indeed the early loop production matters most so that the physical loop number density just dilutes as $1/t^{3\nu}$, as for a gas of particles. 

Since early loop production is most important, the number density in the unstable case may be computed easily noting that at early times $\frac{\left(d_h +t-t'\right)}{d'_h} \rightarrow \frac{2-\nu}{1-\nu} \frac{t}{t'}$. One finds
\begin{equation}\label{eqn:unstabnum}
    n(t) \simeq \xi^{1-p} \frac{a_\star^3 }{t_\star^{3}} \left(\frac{t_\star}{t}\right)^{p-1}
\end{equation}
and we see that the suppression is by a factor $\left(\frac{t_\star}{t}\right)^{p-1}$ with respect to the stable case. Now the physical number density dilutes as $1/t^{3\nu+p-1}$. 

As is clear from Equations \ref{eqn:stabnum} \& \ref{eqn:unstabnum}, our estimate for the total string loop number density depends upon our assumptions about the behavior of the cosmic string network at early times $t_\star$ and at small scales $\xi$. But the conclusion we take from this to use in our benchmark is that there may be a quite large abundance of small cosmic string loops---for a symmetry-breaking scale of order a TeV, the enhancement in the number density of small cosmic strings from $\xi^{1-p}$ may be upwards of $10^{20}$.  

\section{Topological Defect Catalysis} \label{sec:catalysis}

We will discuss the specific example of cosmic string catalysis borne out by our model in Sec \ref{sec:bfstrings}. However, the conceptual physical understanding of the catalysis effect is rather opaque in that case as yet. This is such an exotic effect---in the sense of being far from free field theory---that to build intuition and aid further understanding we begin with exploring similar, more-familiar effects for topological defects of lower dimensionality.

Topological defects are a rich topic in field theory going back all the way to Dirac, who first puzzled over the interactions of charged fermions with magnetic monopoles.\footnote{Of course even Dirac stood on the shoulders of giants and I refer the reader to KT McDonald's manuscript \cite{mcDonald:2015a} for an interesting review of the prehistory.} It is from this setting that we adopt the term `catalysis', which comes from the Callan-Rubakov effect of GUT monopoles catalyzing proton decay. In this example the need for inelastic interactions can be seen elementarily from the far infrared. There has been much progress in the last century of understanding both the infrared and ultraviolet pictures of the electric and magnetic charges.

While it is not usually referred to with this language, the nontrivial intertwining of charged fermions with instantons has also been appreciated since 't Hooft. The physics of these zero-dimensional topological defects is naturally understood in fully Euclidean space, and this makes it easy to see the role and peculiarity of fermion zero modes and the violation of anomalous symmetries. The drawback and disambiguation with the other cases is that instantons do not have a relic abundance and do not truly undergo scattering, so an assertion of `unitarity-limited' interactions is not immediately sensible. But we nonetheless distinguish between some large action cost for producing the instanton, and the fermion interactions which come along with unit probability.

Taking these together, we find it natural to conjecture that all of the phenomena discussed admit some sort of unified description---perhaps in the language of generalized global symmetries---but the analogy has thus far not been made precise. We will below try to call attention to the similarities, but a full exploration is postponed to future work.

At the very least, we can speak about the defects in the general case in topological language.
In this section our scenario of interest is the classical, static Yang-Mills(-Higgs) equations of motion solved on a sphere at the compactification of asymptotic infinity of some number of spatial dimensions $S^k$. The configuration space in this problem for a gauge theory of the group $G$ (possibly spontaneously broken to the subgroup $H$) consists of the gauge boson $A$ (and scalar $\Phi$) fields, which are locally defined maps from this sphere at infinity into the gauge group $S^k \rightarrow G$ (or $S^k \rightarrow G/H$). Such maps are classified by the homotopy group $\pi_k(G)$ (or $\pi_k(G/H) \simeq \pi_{k-1}(H)$) which counts the topologically distinct sectors---a generalized notion of `winding numbers'. These classify the topological defects in a given theory.

\begin{figure}[h]
    \centering
        {\includegraphics[clip, trim=0.0cm 0.0cm 0.0cm 0.0cm, width=0.7\textwidth]{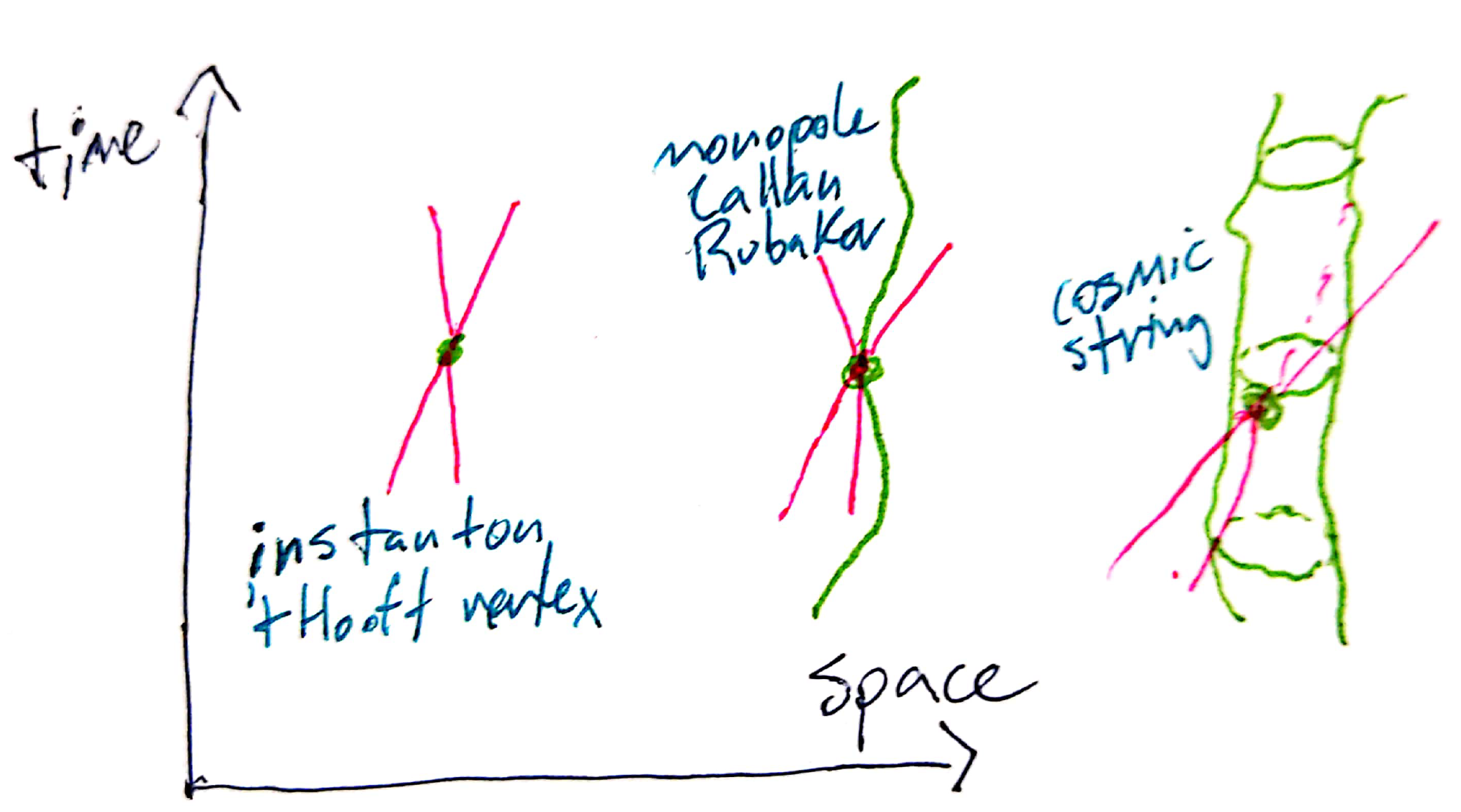}}
    \caption{Cartoon of catalysis processes for topological defects of different dimensions.}
    \label{fig:catalysis}
\end{figure}

\subsection{Instantons and Fermi Zero-Modes} \label{sec:instantons}

Instantons in Yang-Mills theories are dimension-zero topological defects. 
Such solutions exist when there are nontrivial maps from the boundary of Euclideanized spacetime to the gauge group. In the unbroken electroweak case, maps $W: S^3 \rightarrow SU(2)\cong S^3$ are classified by a winding $\bb{Z}$. Instanton configurations are those with nonzero topological charge $\int W \tilde{W}/(16 \pi^2)\neq 0$, and the field non-trivially winds around spacetime. 

\subsubsection{Infrared Inelastic Interactions}
    
The inclusion of fermion fields in the path integral can qualitatively affect the instanton solutions through the presence of fermionic zero-modes. `Zero-modes' refers to zero-energy eigenmodes of the Hamiltonian, and in the bosonic case these are the `collective coordinates' of the object---be they orientation in the physical space or in an internal space as a result of Higgsing. 
However, Pauli-Fermi-Dirac exclusion leads fermionic zero-modes to behave quite differently. With a given  background of the (Euclidean) metric and gauge fields, we follow Fujikawa \cite{Fujikawa:1979ay} and diagonalize the fermions' quadratic action using the Dirac eigenspinors $\varphi_n(x)$ 
\begin{equation}\label{eqn:dirac}
    \gamma^\mu D_\mu \varphi_n = \lambda_n \varphi_n
\end{equation}
where $\gamma^\mu D_\mu \equiv \slashed{D}$ is the Hermitian Dirac operator for that background and $\lambda_n$ are the real eigenvalues. We can write Grassmann-valued Dirac fields $\psi, \bar{\psi}$ in terms of Grassmann coefficients $a_n, \bar b_m$ of these eigenspinors
\begin{equation}
    \psi(x) = \sum_n a_n \varphi_n(x), \qquad \bar\psi(x) = \sum_m \varphi_m^\dagger(x) \bar b_m,
\end{equation}
where the conjugate spinors have the decomposition into $\varphi_m^\dagger(x) \overleftarrow{(\slashed{D})}  =  \lambda_m \varphi_m^\dagger(x)$. We take the eigenspinors to be orthnormalized as $\int_M \varphi_m^\dagger \varphi_n = \delta_{mn}$.  Recall the chirality operator $\gamma_5\equiv i \epsilon^{\mu\nu\rho\sigma} \gamma_\mu \gamma_\nu \gamma_\rho \gamma_\sigma/4!$, and note that since $\lbrace \gamma_5, \gamma_\mu\rbrace = 0$, $(\gamma_5 \varphi_n)$ is also an eigenspinor with eigenvalue $-\lambda_n$. 
These are then linearly independent, as can be seen by taking $\int_M \varphi^\dagger_n \slashed{D} (\gamma_5 \varphi_n)$ and acting with the derivative to either side.

But in the case of a zero-mode this no longer holds as the eigenvalues are the same, so $\gamma_5 \varphi_0 \propto \varphi_0$. Since $\gamma_5^2 = \mathds{1}$, we conclude $\gamma_5 \varphi_0 = \pm \varphi_0$ and the zero-mode is chiral. If the zero-mode satisfies $\gamma_5 \varphi_0 = + \varphi_0$ it is called right-handed and one with $\gamma_5 \varphi_0 = - \varphi_0$ is called left-handed. On the other hand for an eigenvalue $\lambda \neq 0$ there are independent left-handed and right-handed solutions we can construct by projecting onto chiral subspaces  $\varphi_{\pm n} \equiv P_\pm \varphi_n = \half (1 \pm \gamma_5) \varphi_n$.

The Dirac-Yang-Mills action is then
\begin{align}
    -S = \int_M \bar \psi i \slashed{D} \psi &= \sum_{ij} \int_M (\varphi_m^\dagger \bar b_m) i \slashed{D} (a_n \varphi_n) \\
    &= \sum\limits_{\substack{n \text{ with}\\ \lambda_n \neq 0}} i \lambda_n \bar b_n a_n
\end{align}
as a sum over Grassmann coefficients of the Dirac eigenspinors. The zero modes do not appear---there is no action cost to exciting them. The path integral over the fermion fields can be re-expressed in terms of integration over the Grassmann coefficients as
\begin{equation}
    \int \mathcal{D} \bar \psi \mathcal{D} \psi e^{\int_M \bar \psi i \slashed{D} \psi} = \int \left(\prod\limits_{\substack{\text{ nonzero} \\ \text{modes } n}} d\bar b_n da_n
    \right) \int da_0 \ e^{\sum\limits_{n} i \lambda_n \bar b_n a_n }.
\end{equation}
where there is only one Grassmann integral for the zero-mode since it is chiral and has half the number of independent components. Now we recall that for a Grassmann variable $\xi$, integration behaves like differentiation 
\begin{equation}
    \int 1 \ d\xi = 0, \qquad \int \xi \ d\xi = 1.
\end{equation} 
From which can evaluate the nonzero mode integrals directly
\begin{equation}
    \int d\bar b_n d a_n  e^{i\lambda_n \bar b_n a_n} = \int d\bar b_n d a_n (1 +i \lambda_n \bar b_n a_n) = i\lambda_n
\end{equation}
or just recognize that such a functional integral of fermionic fields leads quite generally to $\det(i\slashed{D}) = \prod i \lambda_n$. But if the Dirac operator has a zero-mode, since its Grassmann coefficient does not appear in the action, the path integral above vanishes manifestly. 

In other words, the integration over the Grassmann coefficient of the zero-mode causes any correlation function to vanish unless enough fermion fields are inserted to saturate all the zero-modes. 
So for a given background $\mathcal{I}$ of instantons in which the fermion species $\lbrace \psi_i \rbrace_i$ have zero modes, the only nonzero correlation functions have an insertion of each of these fermion fields,
\begin{equation}
    \left \langle \mathcal{O} \prod\limits_i  \psi_i \right\rangle_{\mathcal{I}} = \int \mathcal{D} \bar \psi \mathcal{D} \psi \mathcal{O} \left(\prod\limits_i  \psi_i\right) e^{-S_{\mathcal{I}}},
\end{equation}
where $\mathcal{O}$ are whatever other operators and $S_{\mathcal{I}}$ is the action in the instanton background. So the functional integral over the fluctuations around an instanton solution \textit{must} be accompanied by a fermion field for each zero-mode. 

Now as for why there should be such zero-modes, there is a celebrated and beautiful result linking anomalous global symmetries to fermi zero-modes of instantons known as the Atiyah-Singer index theorem \cite{atiyah1963index}. This relates the `index' of the theory's Dirac operator---just meaning the number of zero eigenvalues, which is invariant under small deformations of the background---to the topological density of the gauge field.
Defining $J^\mu_{X}$ as a global $U(1)_X$ current in a theory with an ABJ anomaly with the gauge group G with field strength $F$. 
\begin{equation}
    \text{index}(\slashed{D})
     = \int \frac{F \t{F}}{16 \pi^2} = \partial_\mu J^\mu_{X}.
\end{equation}
The number of zero modes of the fermions in an instanton background in this theory are controlled by the mixed anomaly of $U(1)_X$ and $F$, which results in violation of the otherwise-conserved current.

In the infrared, 't Hooft showed that we can interpret the necessity of fermion zero-modes in instanton background as inducing interactions among the fermions that are controlled by the structure of anomalies \cite{tHooft:1976rip,tHooft:1976snw} also \cite{Kiskis:1977vh,Brown:1977bj,Bitar:1977wy,Christ:1979zm}. In the case of the SM electroweak sector, the Higgsing of the symmetry gives the instantons a characteristic size $\rho \sim v_{EW}^{-1}$, and at low energies these may be accounted for by matching onto so-called 't Hooft vertices. We refer to \cite{Morrissey:2005uza} for a lucid discussion.

Such vertices are heavily suppressed in the infrared just because the instantons are rare in the infrared. This differs from the cases of monopoles and cosmic strings, which are not localized in time. So while the occurrence of an instanton is exponentially suppressed, we want to separate this conceptually from the relationship between the defect and charge violation. We suggest that the fact that each instanton \textit{must} come along with a violation of the charge $\Delta Q_X$ has some interpretation as a `unitarity-limited' effect of the fermi zero-modes. 

\subsection{Monopoles and Callan-Rubakov} \label{sec:monopoles}

Monopoles are dimension-1 topological defects in a spontaneously broken gauge theory $G \rightarrow H$ where the quotient can non-trivially wrap the spatial sphere $\pi_2(G/H) \simeq \pi_1(H) \neq 0$. In particular, the SM case is of electromagnetic monopoles of $U(1)$ QED, which has $\pi_1(U(1)) = \bb{Z}$. Many sorts of complementary approaches have been taken toward studying the interactions between electrically charged fermions and magnetic monopoles over the decades. The literature on the topic of monopoles is enormous, and introductions can be found in the classic reviews \cite{Preskill:1984gd,Preskill:1986kp,Rubakov:1988aq} and recent notes \cite{Tong:2018aa}. I will for now resist the temptation to outline and synthesize \textit{all} of this literature, and content myself to present some basic conceptual pictures.

\subsubsection{Infrared Inelastic Interactions} \label{sec:monopoleir}

Starting from the bottom up, an important strain of analysis has been to show directly in the infrared, $U(1)$ gauge theory description that inelastic, helicity-violating interactions must occur. Indeed, the recognition that QED with both electric and magnetic charges leads to novel effects goes back to Dirac \cite{Dirac:1931kp,Dirac:1948um}, who first described the singular magnetic monopole solution which exists in QED (also Tamm \cite{Tamm:1931dda}).
Novel electric-magnetic interplay was found in many guises amenable to elementary arguments \cite{Goldhaber:1965cxe,Zwanziger:1968ams,Peres:1968umt,Lipkin:1969ck,Wu:1976qk} even before attempting a quantum field theoretic treatment \cite{Schwinger:1966nj,Zwanziger:1968rs}.

Historically, it was this endeavor which sparked high energy theorists' enduring love affair with topology \cite{Wu:1975es,Wu:1976ge,Greub:1975mi}. Fibre bundles appear as the natural structure through which to define maps that look locally like functions but have nontrivial global structure, and indeed the fermion wavefunction may be thought of as a `wave-section' of a bundle. But we will here stick to familiar particle theory terminology and defer further formal mathematical description to future work.

\setcounter{footnote}{0}
It all begins with the nonvanishing electromagnetic angular momentum in the presence of an electric charge $e$ and a magnetic charge $g$---even if they are stationary. It is not hard to derive this from the Lagrangian, but we take a shortcut to see this via Schwinger's elementary argument for charge quantization \cite{Schwinger:1969ib} using the Lorentz force.\footnote{The idea for this elementary argument was invented many times independently, with Saha mentioning it in 1936 in the Indian Journal of Physics \cite{Saha:1936mn} and Fierz in 1944 giving it in German in the Swiss journal Helvetica Physica Acta \cite{Fierz:1944zt}. Americans caught on late, with the Physical Review publishing it in a short letter by H. Wilson in 1949 \cite{Wilson:1949zz,Saha:1949np}.}
We imagine a nonrelativistic test particle of mass $m$ and electric charge $e$ moving with velocity $\vec{v}$. In the field of a monopole with magnetic charge $g$, $\vec{B} = g \vec{r}/r^3$, the Lorentz force law is
\begin{equation}
    m \frac{d\vec{v}}{dt} = e \left(\vec{E} + \vec{v} \times \vec{B}\right) = e g \vec{v} \times \frac{\vec{r}}{r^3}.
\end{equation}
Considering the moment of this equation by taking the cross product with the radial vector $\vec{r}$, and with some linear algebra providing $d\hat{r}/dt = \vec{r} \times (\vec{v}\times \vec{r})/r^3$ for the unit vector $\hat{r} \equiv \vec{r}/r$, one finds a modified conserved angular momentum vector  
\begin{equation}
    0 = \frac{d\vec{J}}{dt} = \frac{d}{dt}\left[\vec{r} \times m \vec{v} - e g \hat{r}\right],
\end{equation}
with an extra radial component proportional to both charges! The orbital component of the angular momentum is not quantized, but in light of quantum mechanics we may sensibly postulate the quantization of the perpendicular, radial component of the angular momentum \cite{Mcintosh:1970gg}. If this is forced to take integer or half-integer values, we get Dirac's quantization condition  
\begin{equation}
    e g \in \bb{Z}/2.
\end{equation}
The minimal monopole with $g=1/2e$ produces a magnetic flux through a Gaussian surface of $\Phi_0 = 4 \pi g = 2\pi/e$, matching the flux quantization condition of the minimal Abelian Higgs as we saw in Appendix \ref{sec:ahstrings} above.

In any case, the result that the electromagnetic field configuration stores $\vec{L} = - e g \hat{r}$ persists in the relativistic theory, and immediately poses a puzzle. If we consider shooting a charged fermion at a monopole, $\hat{r}$ pointing from the electric to the magnetic charge would switch sign discontinuously as one moves past the other. 
This necessary change in the sign of the electromagnetic angular momentum as the fermion passes through the monopole means that the scattering boundary conditions must pair incoming fermions of one helicity with outgoing fermions of the other, such that the spin flip makes up for the electromagnetic contribution \cite{Kazama:1976fm,Callan:1983ed}. There \textit{must} be an inelastic interaction, and the purely IR description means that the effect does not vanish as the symmetry-breaking scale is taken to infinity. 

Recent work taking advantage of on-shell techniques has explicitly constructed the $S$-matrix elements describing the scattering of electric and magnetic charges from the infrared \cite{Csaki:2020inw,Csaki:2020yei}.  
This beautifully overcomes the impossibility of a local, Lorentz-invariant Langrangian description as shown by Dirac and Zwanzinger \cite{Dirac:1948um,Zwanziger:1970hk,Zwanziger:1972sx,Weinberg:1965rz}. Here the catalysis effect may simply be stated as the absence of forward scattering between these states, leading to inelastic interactions at the unitarity limit. With no forward scattering present, note that we can speak only of the $S$-matrix element and not of an amplitude, whose definition assumes a decomposition heuristically as $S_{ab} = \delta_{ab} + \mathcal{M}_{ab}$. This unusual structure is also closely related to the violation of crossing symmetry of electric-magnetic scattering. Physically this results from the fact that a configuration with an incoming electric and magnetic charge has an angular momentum contribution from the electromagnetic field that is entirely absent if one of the initial state particles is crossed to an outgoing antiparticle.

\subsubsection{Into the Defect} \label{sec:monopoleuv}

While the effect may be seen purely in the infrared theory, further insight may be gained from how this appears out of the microphysics. Once the UV completion of Dirac's monopole had been found from Higgsed non-Abelian gauge theory by 't Hooft and Polyakov \cite{tHooft:1974kcl,Polyakov:1974ek}(see also \cite{Arafune:1974uy,Prasad:1975kr} and dyons shortly thereafter \cite{Julia:1975ff,Hasenfratz:1976vb,Tomboulis:1975qt,Callan:1975yy,Boulware:1976tv}), one could delve into the monopole to attempt to understand the microphysics of scattering.
With the full solution in hand, one can observe that the monopole, as a map $\varphi^a(x)$ from physical space to internal space, breaks the separate rotational and gauge orientational symmetries. In the monopole background, the correct conserved angular momentum-like quantum number is the unbroken sum, which sums the contributions
\begin{equation}
\vec{J} = \vec{L} + \vec{T},
\end{equation}
where the normal angular momentum is combined with the generator $\vec{T}$ of a broken $SU(2)$ subgroup of the gauge group, leading to a sort of mixing of internal and external `spin' factors \cite{Jackiw:1975fn,Hasenfratz:1976gr,Goldhaber:1976dp,Goldhaber:1977xw}. This can easily be seen to imply that the chirality-flipping boundary conditions should also flip the $\vec{T}$ quantum numbers \cite{Sen:1984qe}.

From the ultraviolet we can also see that the monopole has a dyonic (or `phase' or `rotor' or `rotator') mode---an electrically-charged degree of freedom living on the monopole core---resulting from the orientation of the Higgs in the internal space. It arises physically because the ``little group'' of the Higgs is parametrized by the coset of the $U(1)$ within the broken gauge group. The importance of this mode in scattering off monopoles was powerfully emphasized by Polchinski \cite{Polchinski:1984uw} who demonstrated that the essential physics can be seen just in the quantum mechanics of $2d$ Weyl fermions interacting with this mode. 
Recently,
Brennan has thoroughly explored the Callan-Rubakov physics from this perspective through explicit calculation \cite{Brennan:2021ewu,Brennan:2021ucy}, following earlier work including \cite{Isler:1987xn,Affleck:1993np}. 
The dyonic degree of freedom living in the monopole core is identified as arising from the longitudinal mode of the broken gauge bosons. 

The physics in the spherically-symmetric sector simplifies to two dimensions, and an incoming fermion can be seen to undergo inelastic scattering by interacting with this mode. If we integrate it out, we can derive the boundary conditions for fermions in the $1+1$-dim theory of $s$-wave scattering. And indeed, Brennan shows explicitly that the boundary conditions violate those global symmetries that have an ABJ anomaly with the broken gauge group.
Heuristically, the intuition is that exciting the dyonic mode turns on an electric field which activates the ABJ anomaly with, in the Callan-Rubakov case, $B+N_cL$ as
\begin{equation}
    \partial_\mu J^\mu_{B+N_cL} \propto \int F \tilde{F} \propto \int \vec{E}\cdot\vec{B},
\end{equation}
where these are the familiar electric and magnetic fields of electromagnetism, and this leads to violation of $B+N_cL$. The Callan-Rubakov effect is then that a monopole passing through a proton, by means of interacting with the dyonic mode, realizes this violation by turning the proton into a positron with unitarity-limited probability \cite{Rubakov:1982fp}.

To gain further conceptual understanding it is useful to bosonize the physics of 2d fermions in the $(r,t)$ half-plane \cite{Callan:1982ac,Callan:1982au,Callan:1982ah,Callan:1983tm,Dawson:1983cm} in the spirit of the original duality between the scalar `sine-Gordon' theory and the `Thirring model' of interacting fermions \cite{Coleman:1974bu,Mandelstam:1975hb}. See also \cite{Rubakov:1988aq} for an early review. Heuristically, one realizes each fermion flavor bilinear as a compact bosonic degree of freedom $\bar \psi_i \psi_i \sim e^{i\phi_i}$. Now this is still perturbatively a theory of $2d$ scattering, but we can actually solve the bosonic theory to find the ground states. The bosonized fields have the interpretation that $\phi_i(r=0)$ is the $i$th particle number on the monopole core. 
One has a Lagrangian of the form 
\begin{equation} \label{eqn:bosonizedlag}
    \mathcal{L}_\text{int} \sim \sum\limits_{\text{species} \ i} \mu_i^2 \cos(2\pi \phi_i) + \sum\limits_{\substack{\text{unbroken}\\ U(1)_k}} \frac{g_k^2}{r^2} \left(\sum\limits_{i} q_i \phi_i\right).
\end{equation}
In the first term $\mu_i$ is related to the fermion mass, and the periodic potential imposes that the theory has vacua with different numbers of fermions lodged on the core. The second `Coulomb' term, which includes sums over $U(1)$ subgroups of unbroken non-Abelian symmetries, enforces these symmetries such that only neutral combinations of fermions appear together on the core.

In the SU(5) grand unified case, one finds that the monopole core contains an integer multiple of the combination $uude$, which is gauge invariant and furthermore has $B-N_cL=0$, but $B+N_cL = -2N_c$. Since transitions between the vacua have finite action, the ground state of the monopole is a superposition $\sum \left| n \right\rangle$ with indefinite baryon plus lepton number. So when the monopole interacts with the external world, it plainly violates $B+N_cL$ \cite{Callan:1982ac,Csaki:2021ozp}. We expect that transitions between degenerate vacua with different numbers of fermions in this reduced model correspond to fermi zero modes of the monopole in the full 4d theory, as a sort of parallel to the instanton case.

Finally, we note there are a variety of exotic effects arising from the inclusion of fermions---the basic fact being that fermion number may fail to be a good symmetry in the vicinity of a monopole. Jackiw \& Rebbi \cite{Jackiw:1975fn} suggest that a monopole with a fermi zero-mode 
has fractional fermion number $\half$, and Goldstone \& Wilczek \cite{Goldstone:1981kk} made this a bit more intuitive by studying kinks in a $1+1$d model of polyacetylene and showing the same effect. Goldhaber \cite{Goldhaber:1976dp} resolved its seeming violation of spin-statistics \cite{Hasenfratz:1976gr,Hasenfratz:1976vb} by careful analysis (see also \cite{Jackiw:1975fn,Rossi:1977im,Callias:1977kg,Callias:1977cc,Blaer:1981ui,Wilczek:1981dr,Yamagishi:1982wp,Matveev:1988pj}). 

\subsection{Cosmic Strings} \label{sec:bfstrings}

The strings we studied above are an example of dimension-2 topological defects in a spontaneously broken gauge theory $G \rightarrow H$ where the quotient can non-trivially wrap the transverse circle $\pi_1(G/H) \simeq \pi_0(H) \neq 0$ \cite{Schwarz:1982ec}. $\pi_0(H)$ is simply the number of connected components of the infrared symmetry group. For reviews see e.g. \cite{Hindmarsh:1994re,Vilenkin:2000jqa,Copeland:2009ga}.

Now in contrast to the prior cases, the scattering of fermions off cosmic strings has seen much sparser study. 
Though aspects of the interactions of fermions with $2+1$d Abelian Higgs Nielsen-Olsen \cite{Nielsen:1973cs,deVega:1976xbp} vortices have been studied beginning with \cite{Nohl:1975jg,deVega:1976rt,Jackiw:1981ee,Weinberg:1981eu}. 
For cosmic strings, the literature begins with \cite{Alford:1989ie,Alford:1988sj} which we will use as our model of the scattering in which charged fermions interact with a scalar condensate in the string. Other related work on fermions and cosmic strings includes \cite{Brandenberger:1991dr,Hill:1987qk,Ganoulis:1989hz,deSousaGerbert:1988kpn,Semenoff:1987ki,Perkins:1990uv,Widrow:1988rc,Hill:1986ts}. 

The string case has only been studied from a UV perspective for specific models of the string core, so in this section we will begin there. We will write down our own microphysical model which maps to the previously-studied case of \cite{Alford:1989ie} such that we may appropriate their results. We note, however, that our understanding of the ultraviolet picture here is far less well-developed than in the monopole case. Afterwards, we'll offer some comments about the infrared picture, but a full analysis is postponed to future work.

\subsubsection{Into the Defect} \label{sec:stringuv}

An explicit calculation of catalyzed baryon number violation by a cosmic string was performed by Alford, March-Russell, and Wilczek \cite{Alford:1989ie}. They have cosmic strings of some gauged $\widetilde{U(1)} \rightarrow \bb{Z}_N$, and they introduce a scalar $\phi$ which condenses on the cosmic string and couples to a lepton and a quark as 
\begin{equation}
    \mathcal{L} \supset \lambda \left(\phi \bar \psi_q \psi_\ell + \phi^\star \bar \psi_\ell \psi_q \right).
\end{equation}
So $\phi$ is a genuine leptoquark carrying color and electromagnetic charge and $B+N_cL$, and its condensation breaks all these symmetries. It should also provide bosonic zero modes which can carry off the imbalance of charges when the string catalyzes $p^+ \rightarrow e^+$, as with the monopole core. They compute that an incoming quark wavefunction is amplified towards the core of the string and undergoes an inelastic interaction at the unitarity limit
\begin{equation}\label{eqn:xsecinel}
    \frac{d\sigma}{d\theta} \sim \frac{1}{p} \sin^2(\pi \alpha) \left(\frac{p}{v}\right)^{4\left|\alpha-\half\right|},
\end{equation}
where I am generalizing Eqns 18-21 of \cite{Alford:1989ie} and have $\alpha = k q / N$, with $k$ the winding number of the string, $q$ the $\widetilde{U(1)}$ charge of the incoming quark, and $N$ the order of the infrared discrete symmetry. We recognize this form from the elastic Aharonov-Bohm scattering, but now with additional suppression by ultraviolet scales except for the case $\alpha = 1/2$. Then it is only configurations with maximal BF charge, which for $k=1$ is $q = N/2$, for which the scattering is entirely unsuppressed by ultraviolet scales as in the Callan-Rubakov effect. We note that while their string superconducts in the  electromagnetic sense, the SM gauge fields are explicitly ignored and manifestly not required for the catalysis. 

Now our purpose here is slightly different---in particular we wish to preserve the SM selection rule which stabilizes the proton, so we instead add a gauge-singlet scalar $\chi$ with solely $B+N_cL$ charge equal to $2 N_c \gen$. This condenses on the core of the string and breaks the classical global $U(1)_{B+N_cL} \rightarrow \bb{Z}^{B+N_cL}_{2N_c \gen}$, down to the exact global symmetry in the SM. We must include a broader scalar sector to communicate the $U(1)_{B+N_cL}$-breaking to the SM fermions while respecting the unbroken symmetries. As our benchmark example, we take the scalars listed in Table \ref{tab:newcharges} and the mediation is ultimately through the leptoquark and diquark $\omega_\ell, \omega_q$ both with hypercharge $+8$ and $B-N_cL$ charge $-2$ but different global charges $B+N_cL = +4, -2$, see Figure \ref{fig:feynDiag}. We imagine the mediators have $v$-scale masses from $\left\langle \Phi\right\rangle$, so integrating out these mediators around the vacuum generates an operator
\begin{equation} \label{eqn:highdimop}
    \mathcal{L} \sim \frac{1}{v^{15}} \chi \left(\bar u \bar u \bar d \bar e\right)^3 + \text{h.c.}
\end{equation}
The situation of cosmological interest is of three protons incident on a cosmic string, so we interpret this by shuffling the quarks and leptons into spinors of three protons and three electrons,
\begin{equation} \label{eqn:alfordAnalogue}
   \mathcal{L} \sim \lambda \left(\chi \bar \psi_{3p} \psi_{3e} + \chi^\star \bar \psi_{3e} \psi_{3p}\right),
\end{equation}
where we are ignoring many subtleties.
This equation is really just a heuristic for us, the importance of which is to evince the boundary condition at the string core for an incoming charged fermion. We aren't actually interested in the operator of Equation \ref{eqn:highdimop} near the vacuum; the important interaction with $\chi$ will take place in the core of the string, where we assume the mediating fields $\omega_i$ are light and so do not heavily suppress the interaction. 

In the analogue of this setup in \cite{Alford:1989ie}, Alford, March-Russell, and Wilczek computed the two-dimensional quantum mechanics problem of an incoming $\psi_{3p}$ particle incident on the cosmic string and scattering inelastically into an outgoing $\psi_{3e}$. Now clearly the three-proton state $\psi_{3p} = {}^3 \text{Li}$ is not a stable bound state, but in the cosmological case of interest lumps of three protons are bound in the ${}^7 \text{Li}$ nuclear wells. It is these lithium nuclei which are incident on the string, but without further apology we assume that we can just use the calculation for an incident $\psi_{3p}$ and ignore the neutron background they live in.

They set out to solve the Aharonov-Bohm scattering problem with an incident plane wave and an outgoing scattering wave
\begin{equation}
    \lim\limits_{\rho\rightarrow\infty} 
    \begin{pmatrix}
    \psi_{3p} \\
    \psi_{3e}
    \end{pmatrix} =
    \begin{pmatrix}
    \psi^{i}_{3p} \\
    0
    \end{pmatrix} + 
    \begin{pmatrix}
    \psi^{s}_{3p} \\
    \psi^{s}_{3e}
    \end{pmatrix}
\end{equation}
where the fermions obey equations of motion in the string background which in our case have approximately the form
\begin{equation}
    \begin{pmatrix}
    \slashed{\partial} - i q g \slashed{A} & \lambda \langle \chi \rangle^\star \\
    \lambda \langle \chi \rangle & \slashed{\partial} - i q g \slashed{A}
    \end{pmatrix}
    \begin{pmatrix}
    \psi_{3p} \\
    \psi_{3e}
    \end{pmatrix} = 0,
\end{equation}
where $A$ here is the $U(1)_{B-N_cL}$ gauge boson, g the gauge coupling, and $q=N_c \gen$ the $B-N_cL$ charge of the incoming three proton state. $\slashed{A}$ is the discrete Aharonov-Bohm potential at long distances as in Equations \ref{eqn:stringPot}, \ref{eqn:anovortex}, and $\langle \chi \rangle$ vanishes except in the string core. The wavefunctions in and outside the core must be carefully matched, and the cross-section is proportional to the magnitude of the outgoing three-positron current. 

Then by mapping our case to that studied in \cite{Alford:1989ie}, we may appropriate the cross section they computed, Equation \ref{eqn:xsecinel}, with $N = 2N_c \gen$ the infrared discrete subgroup of $B-N_cL$ under which three protons have charge $q = N_c \gen$. By the fact that we've chosen $\chi$ to preserve the SM proton stability selection rule, we have unsuppressed inelastic scattering in our case of $\psi_{3p} \rightarrow \psi_{3e}$. 
That is, our choice of leptoquarks and Higgs charges has picked out exactly the process of interest
\begin{equation}
    \frac{\sigma}{\ell}\left( 3 p^+ + \text{string} \rightarrow 3 e^+ + \text{string}\right) \sim \Lambda_{QCD}^{-1}.
\end{equation}
where $\frac{\sigma}{\ell}$ is the cross-section per unit length, as computed for a long string. We note that as we saw above with monopoles, crossing symmetry may be violated for topological defects, and in particular for us \cite{Alford:1989ie} show the cross-section behaves differently for bosons, as if we crossed one of the initial protons to a final antiproton.

Let me make a couple further comments on the properties of this model. Firstly, the condensation of $\chi$ on the string core allows it to superconduct bosonic global $B+N_cL$ currents, as first pointed out by Witten \cite{Witten:1984eb}. Such modes presumably carry off the  $U(1)_{B+N_cL}$ charge which appears from the IR to be violated by $2 N_c \gen = 0 \text{ (mod } 2N_c \gen)$ units, but this has not been studied in detail. This behavior would parallel the Callan-Rubakov effect, where $2N_c$ units of $B+N_cL$ charge are deposited onto the monopole core. 


Secondly, the $B-N_cL$ Higgs $\Phi$ also admits low-energy many-fermion interactions with the SM fermions, for example 
\begin{equation}
    \mathcal{L} \supset \frac{1}{v^{15}} \Phi (QL\bar u^\dagger \bar d^\dagger)^3,
\end{equation}
which could be implemented with additional leptoquarks. With some similar arrangement of these products of fundamental fermions into Dirac fermions, this could imply that the string has \textit{fermionic} zero-modes of the sort that were explicitly constructed by Callan \& Harvey \cite{Callan:1984sa}. But here the fact that such collections of fermions are not good bound states would seem to pose a larger problem, and it is at least nonobvious whether there could really be electromagnetically-charged multi-fermion zero modes on the string. It is not clear what role there could be for these interactions, if any.

\subsubsection{Infrared Inelastic Interactions?}

Our explicit example above is an ultraviolet description, and despite it being speculated long ago that this feature should be quite general, it has not had the sort of intense study that has informed our understanding of the Callan-Rubakov effect. Agreeing with the speculation that this is a feature of infrared $\bb{Z}_N$ gauge theory, it should have a natural description in the dual BF theory. 

When we studied the $\bb{Z}_N$ BF theory above we studied operators which create probe particles or strings, and we found that they have a long-range topological interaction. The $\bb{Z}_N$ symmetry is there reflected in the periodicity of the Aharonov-Bohm phases from the long-range interaction---we can only detect charges $(\text{mod } N)$. But then we should also be able to create charges in $N$-fold batches, since these are collectively uncharged. However, the BF theory does not include any such short-range interactions.

Such an interaction of line or surface operators should be described by operators which need not be integrated over a closed loop, but rather have a boundary where the operators meet to interact. Then this must be a line operator which is gauge-invariant, and we came across just such an object above---the combination $(\partial_\mu \phi - N A_\mu)$---which when exponentiated indeed provides the further line operator
\begin{equation}
     e^{i \int_{\gamma_p} (\partial_\mu \phi - N A_\mu) dx^\mu} \rightarrow e^{i \left[\phi(p)- N \int_{\gamma_p} A_\mu dx^\mu\right] },
\end{equation}
where $\gamma_p$ is a line from infinity to the point $p$. 
This is a sort of 't Hooft operator for the two-form $B$, but comes along with $N$ Wilson lines as a result of the BF interaction. In general these lines need not be coincident, and they can describe e.g. a process where $N/2$ charge-1 fermions have a point interaction with the Goldstone and scatter inelastically into another $N/2$ fermions. And of course there is a similar operator describing $N$ strings meeting along a line.

However, in the pure BF theory these 't Hooft operators are trivial in that they vanish on the support of the equations of motion. 
It is clear that one missing ingredient is light charged fermions, but their addition is nontrivial.  We plan to pursue this further in future work.

\bibliographystyle{jhep}
\bibliography{lithium}

\end{document}